\documentclass[12pt]{iopart}
\usepackage{hyperref}
\usepackage{iopams}  
\usepackage{dsfont}

\makeatletter
\renewcommand{\@fnsymbol}[1]{%
  \ifcase#1
    \or \S            % 1st footnote → §
    \or *             % 2nd
    \or \dagger       % 3rd
    \or \ddagger      % 4th (change if you dislike this too)
    \or \mathsection  % 5th, etc. (or anything you want)
  \else
    \@ctrerr
  \fi}
\makeatother

\usepackage[font=footnotesize,labelfont=bf]{caption}

\usepackage{url}
\usepackage{bm,bbm}
\usepackage{verbatim}
\expandafter\let\csname equation*\endcsname\relax

\expandafter\let\csname endequation*\endcsname\relax
\usepackage{amsmath}

\usepackage[capitalise]{cleveref}
\usepackage{dsfont}
\usepackage{lipsum}
\usepackage{xcolor}
\usepackage{soul}
\usepackage{caption}
\usepackage{subcaption}
\usepackage{graphicx}

\usepackage{cite}
\usepackage{amsthm,amsmath}

\usepackage{booktabs}
\usepackage{algorithmicx}
\usepackage{algpseudocode}
\algrenewcommand\algorithmicrequire{\textbf{Input:}}
\algrenewcommand\algorithmicensure{\textbf{Output:}}

\theoremstyle{definition}

\newtheorem{ex}{Example}

\newcommand{\de}{\mathop{}\!\mathrm{d}}

\usepackage{titlesec}
\titleformat{\paragraph}
  {\normalfont\normalsize\bfseries}{\thesubsubsection.\arabic{paragraph}}{1em}{}

\newcommand{\R}{\mathbb{R}}

\newcommand{\argmax}{\operatorname{argmax}}

\makeatletter
\def\@mkboth#1#2{}
\newlength\appendixwidth
\preto\appendix{\addtocontents{toc}{\protect\patchl@section}}
\newcommand{\patchl@section}{%
  \settowidth{\appendixwidth}{\textbf{Appendix }}%
  \addtolength{\appendixwidth}{1.5em}%
  \patchcmd{\l@section}{1.5em}{\appendixwidth}{}{}%
}
\makeatother

\begin{document}

\title{The Most Dispersed Subset of Random Points in $\R^d$}

\author[1]{Fabio Deelan Cunden$^{1,4}$, Noemi Cuppone$^2$, Giovanni Gramegna$^{3,4}$, Pierpaolo Vivo$^{2}$\footnote{Corresponding author. Email: pierpaolo.vivo@kcl.ac.uk}}

\address{$^1$Dipartimento di Matematica, Universit\`a degli Studi di Bari, I-70125 Bari, Italy}
\address{$^2$Department of Mathematics, King’s College London, Strand, London, WC2R 2LS, United Kingdom}
\address{$^3$Dipartimento di Fisica, Universit\`a degli Studi di Bari, I-70126 Bari, Italy}
\address{$^4$INFN, Sezione di Bari, I-70126 Bari, Italy}

\begin{abstract}
Consider a population of $N$ individuals, each having $d\geq 1$ different traits, and an additive measure, called \emph{dispersion}, which rewards large pairwise separations between traits. The goal is to select $M\leq N$ individuals such that their traits are as dispersed as possible. We compute analytically the full statistics (including large deviation tails) of the maximally achievable dispersion among sub-populations of size $M$ when the traits are independent and identically distributed. Two complementary approaches are developed, one based on a mean-field theory for order statistics, and the other on the replica method from the field of disordered systems. In all dimensions $d$, and for rotationally symmetric distributions, the optimal subset for large populations consists of all points lying outside a $d$-dimensional ball whose radius is determined self-consistently. For a single trait ($d=1$), the statistics of the maximal dispersion can be tackled for finite $N,M$ as well. The formulae we obtained are corroborated by numerical simulations on small instances and by heuristic algorithms that find near-optimal solutions.
   
\end{abstract}

%\tableofcontents
\newpage

\section{Introduction}

Selecting a subset of size $M\leq N$ from a larger population of $N$ individuals, which maximises a notion of ``dispersion'' among their traits, is a fundamental problem in numerous disciplines. 

For instance, the accuracy of polls and surveys often relies on picking a sufficiently representative sample of individuals that cover a broad spectrum of political opinions, personal or health characteristics \cite{health,twilight}. Policies are often drawn to ensure that committees, panels, juries, university cohorts, and several collective bodies are hired or nominated ensuring some diversity of viewpoints, backgrounds, and other attributes \cite{committee1,committee2,committee3,committee4}. In facility location problems, one needs to distribute facilities to serve different geographical regions while minimising interference \cite{facility1,Erkut1989}. 
In finance, portfolio diversification strategies aim to select assets that are minimally correlated to reduce overall risk \cite{portfolio}. Other contexts to which the maximum dispersion problem may be applicable include selection of diverse traits in plant and animals controlled breeding \cite{genetic1,genetic2} and workforce management \cite{workforce}. The abundance of different ``traits'' in a model of integer vectors was shown to correlate with the appearance of Zipf's law \cite{corberi}. 

In the Operations Research field, these questions fall generally under the category of \emph{Maximum Diversity/Dispersion Problems} (MDP) (see \cite{review1} for a recent review). 
For the most common measures of dispersion and for generic number of traits, the MDP is known to be NP-hard, placing it among a prominent class of computationally challenging combinatorial optimisation problems~\cite{Ghosh1996,Kuo,Fernandez} that are usually tackled using heuristic and metaheuristic algorithms that may find efficient and accurate solutions for large instances~\cite{QKAlg,gloverHeuristics, Marti2013}. 

Populations with randomly assigned traits constitute important validation benchmarks for the performance of these heuristic algorithms (see e.g. the standard benchmark library MDPLIB, which includes many instances with individuals' traits sampled uniformly at random~\cite{Kuo}). However, analytical results on the statistics of the maximally achievable dispersion and the ``geometry'' of the maximising subset for random traits are surprisingly scarce. 

\textcolor{black}{A natural question is why uniform random sampling of  $M$ individuals does not suffice. While unbiased, a uniformly drawn subsample concentrates around the high-probability regions of the distribution by the law of large numbers, thereby under-representing rare (low-probability) traits.
Maximising dispersion, by contrast, prioritises configurations with large pairwise distances and thus actively promotes the inclusion of the tails, leading to a more complete coverage of the trait space. This distinction between \emph{unbiasedness} and \emph{coverage} may be quite important in practical applications.}

In this paper, we address a specific instance of MDP -- closely related to the so-called \emph{quadratic Knapsack Problem} \cite{QKP} -- on populations having \emph{random} traits in a $d$-dimensional space. This is based on the \emph{$M$-dispersion} measure (see Section \ref{sec:problemstatement} for details), which rewards large \textcolor{black}{aggregate} pairwise separations between traits. We tackle this problem in all dimensions $d\geq 1$ for large $N,M$ with their ratio fixed using two complementary approaches: one based on a mean-field theory for order statistics, and one based on the replica method of disordered systems. Moreover, for $d=1$ and $N,M$ finite, we show that -- while exact formulae are very challenging to obtain -- an approach based on finite-$N,M$ approximants gives asymptotically exact results that match the numerics on small instances with striking precision. For one of the very few available theoretical studies on quadratic Knapsack Problem with random instances, see \cite{randomQKP}.

For large $N,M$ and in all dimensions $d \geq 1$, we compute the full distribution of the maximal dispersion in the sense of large deviations. From this distribution, all cumulants can be obtained, to leading order, by differentiation in the large-$N,M$ limit. As a consequence, the probabilities of rare events -- away from the typical Gaussian regime -- in which the maximal dispersion is anomalously larger or smaller than its expected value are characterised exactly for large $N,M$. \textcolor{black}{Moreover, we find that a subset whose dispersion is asymptotically close to the maximal one is composed of individuals in the $d$-dimensional trait space lying outside a ball, whose radius can be determined explicitly.} Finally, the analytical results are in excellent agreement with numerical simulations for moderate instance sizes and with heuristic solutions for larger problems.

%%%
\subsection{Outline of the paper}
The plan of the paper is as follows. In Section \ref{sec:problemstatement}, we formulate the problem and establish our notation. Section \ref{sec:d1} is devoted to the $d=1$ (single trait) case: in subsection \ref{sec:order_stat} we discuss the geometry of the optimising subset; in \ref{sec:meanfieldD1} we formulate the mean-field theory for the maximal $M$-dispersion in the large-$N,M$ limit, while in \ref{sec:backfiniteN} we discuss balanced configurations and the finite-$N,M$ approximants of the maximal $M$-dispersion. In Sec. \ref{sec:continuous} we extend the mean-field approach to $d\geq 1$, evaluating separately the typical maximal $M$-dispersion \ref{subsec:typical} and the Scaled Cumulant Generating Function (SCGF) and rate function in real space \ref{subsec:SCGF}. The same result for the SCGF is recovered in Sec. \ref{sec:replica} from the replica method. Our results in dimension $d>1$ are tested numerically using a heuristic greedy algorithm, which we describe in Section~\ref{sec:heuristics}.  Section~\ref{sec:conclusion} is devoted to concluding remarks and outlook for future work. The Appendices cover technical definitions and results.

\section{Problem setting}\label{sec:problemstatement}
\subsection{$M$-Dispersion function}
Given a set of $N$ points $\bm x_i\in\R^d$ and a binary vector $\sigma = (\sigma_1,\dots,\sigma_N)^t \in\{0,1\}^N$, such that $\sum_{i=1}^N \sigma_i = M$,  we define the $M$-dispersion  as
\begin{align}
    D^{M}(\bm x_1,\dots,\bm x_N|\sigma) :&=\sum_{i,j=1}^N |\bm x_i-\bm x_j|^2\sigma_{i}\sigma_{j}\ . \label{DefMDispersion}
\end{align}
We indicate by $|\bm x|=\left(\sum_{i=1}^d x_i^2\right)^{\frac{1}{2}}$ the Euclidean norm, and the binary vector $\sigma\in\{0,1\}^N$ encodes which $M$ points are selected: $\sigma_i=1$ if $\bm x_i$ is chosen, and $\sigma_i=0$ otherwise. We define the maximal $M$-dispersion  $D^{M}_{\mathrm{max}}$ as
\begin{equation}
\label{eq:def_Dmax}
    D_{\max}^M(\bm x_1,\dots,\bm x_N):=\max_{\sigma\in\Sigma_{N,M}} D^M(\bm x_1,\dots,\bm x_N|\sigma)\ ,
\end{equation}
with $\Sigma_{N,M}=\{\sigma\in\{0,1\}^N:\sum_{i=1}^N\sigma_i=M\}$.

\textcolor{black}{We note that the additive measure \eqref{DefMDispersion} is one among several possible definitions of dispersion used in the Operations Research literature. Other choices may be better suited to specific applications.} 

\subsection{Statistics of maximal $M$-Dispersion and Scaled Cumulant Generating Function}
Let $\bm X_1,\dots,\bm X_N$ be i.i.d.\ in $\R^d$ with distribution $F$. \textcolor{black}{Since \eqref{DefMDispersion} is a  sum of $M(M-1)$ nonzero terms, and each of them $|\bm X_i-\bm X_j|^2$ is $O(1)$ for distributions with finite second moment, the maximal $M$-dispersion $D_{\max}^M(\bm X_1,\dots,\bm X_N)$ is a random variable of order $O(N^2)$, whenever $M$ is of order $N$.} We therefore consider its rescaled form
\begin{equation}
\frac1{N^2}D^M_{\max}\ , 
\end{equation}
for $M=\alpha N$, and denote with $\kappa_\ell^{(N,M,d)}$ the $\ell$-th cumulant of this rescaled variable. Generically $\kappa_\ell^{(N,M,d)}=O(N^{1-\ell})$, and we define the scaled limits
\begin{equation}\label{eq:kappalimit}
\kappa_\ell^{(d)}(\alpha) := \lim_{\substack{M,N\to\infty \\ M/N=\alpha}} N^{\ell-1}\kappa_\ell^{(N,M,d)}\ .
\end{equation}
The large-$N$ Scaled Cumulant Generating Function (SCGF) is
\begin{equation}
\Phi_\alpha(p) := \lim_{N\to\infty} -\frac1{N}\log \mathbb{E}\big[\e^{-p D^M_{\max}/N}\big]\ ,
\end{equation}
so that 
\begin{equation}
(-1)^{\ell-1}\frac{\de^\ell}{\de p^\ell}\Phi_\alpha(p)\Big|_{p=0} = \kappa_\ell^{(d)}(\alpha)\ .\label{eq:leadingcumulantsfromPhi}
\end{equation}
\textcolor{black}{Under the essentially smoothness assumption of the G\"artner--Ellis theorem~\cite[Theorem II.2]{Ellis1984}, the Legendre-Fenchel transform of $\Phi_\alpha(p)$} gives the large-deviation rate function $\Psi_\alpha(x)$:
\begin{equation}
\Pr\Big(D_{\max}^M = N^2x\Big)\approx \e^{-N\Psi_\alpha(x)}\ ,
\end{equation}
where $\approx$ stands for equality on a logarithmic scale in the $N\to\infty$ limit. The precise relation between the rate function and the SCGF is~\cite{touchette} 
\begin{equation}
    \Psi_\alpha (x)=\sup_p \left(\Phi_\alpha(p)-px\right),\qquad 
        \Phi_\alpha (p)=\inf_x \left(\Psi_\alpha(x)+px\right)\ .\label{eq:legendre}
\end{equation}
We assume that the probability distribution $F$ on $\mathbb{R}^d$ has a rotationally invariant density,
\begin{equation}
    f(\bm x)\de\bm x=r^{d-1}g(r)\de r\frac{\de\hat{\bm n}}{|\mathbb{S}^{d-1}|}\ ,\label{eq:fbmxrotinvdensity}
\end{equation}
where $r=|\bm x|$, and $\hat{\bm n}=\bm x/|\bm x|$ is a vector on the unit sphere $\mathbb{S}^{d-1}$, whose  surface measure is
\begin{equation}
|\mathbb{S}^{d-1}|=\int_{\mathbb{S}^{d-1}}\de\hat{\bm n}=2\pi^{d/2}/\Gamma(d/2)\ . 
\end{equation}
The mean of $f$ is zero; this entails no loss of generality, since the dispersion is invariant under translations. In the next section, we tackle the case of a single trait $(d=1)$, with the trait values of $N$ individuals being random real variables, and the goal is to compute the statistics of the maximal $M$-dispersion and the geometry of the optimising $M$-subset.

\section{Single trait ($d=1$)}\label{sec:d1}

In this section, we first focus on the case of a single trait $d=1$, where the $N$ individual values lie on the real line. The natural ordering of real numbers has an interesting consequence on the ``geometry'' of the subset of $M$ individuals that maximises the $M$-dispersion, as detailed in the next subsection.

\subsection{Geometry of the optimal subset}
\label{sec:order_stat}

In $d=1$, we can characterise the geometry of the optimal $M$-subset that will give maximal dispersion. \textcolor{black}{Assume that the $N$ points are sorted in non-decreasing order, $x_1\leq x_2\leq\cdots\leq x_N$. Then, for each $M\geq 2$,} there exists an optimal $\sigma^{N,M}$ of the form
\begin{equation}
\sigma^{N,M} = (\underbrace{1,\dots,1}_{k},\underbrace{0,\dots,0}_{N-M},\underbrace{1,\dots,1}_{M-k})\ ,
\end{equation}
i.e.\ the $k$ leftmost and $M-k$ rightmost points, for some $1\leq k\leq M$. We call this structure \emph{prefix-suffix}. To prove this, let $\{i_1<\cdots<i_M\}$ be selected indices.  
Write
\[
D^M(x_{i_1},\dots,x_{i_M}) 
= 2(x_{i_1}-x_{i_M})^2 
+2\!\sum_{j=2}^{M-1}\!\big[(x_{i_1}-x_{i_j})^2+(x_{i_j}-x_{i_M})^2\big]
+ D^M(x_{i_2},\dots,x_{i_{M-1}})\ .
\]
The first term is maximised by $i_1=1$, $i_M=N$.  
If there are gaps in the interior indices, shifting them outward increases the sum.  
Thus optimal sets are prefix–suffix blocks.

\begin{figure}
    \centering
    \includegraphics[width=\linewidth]{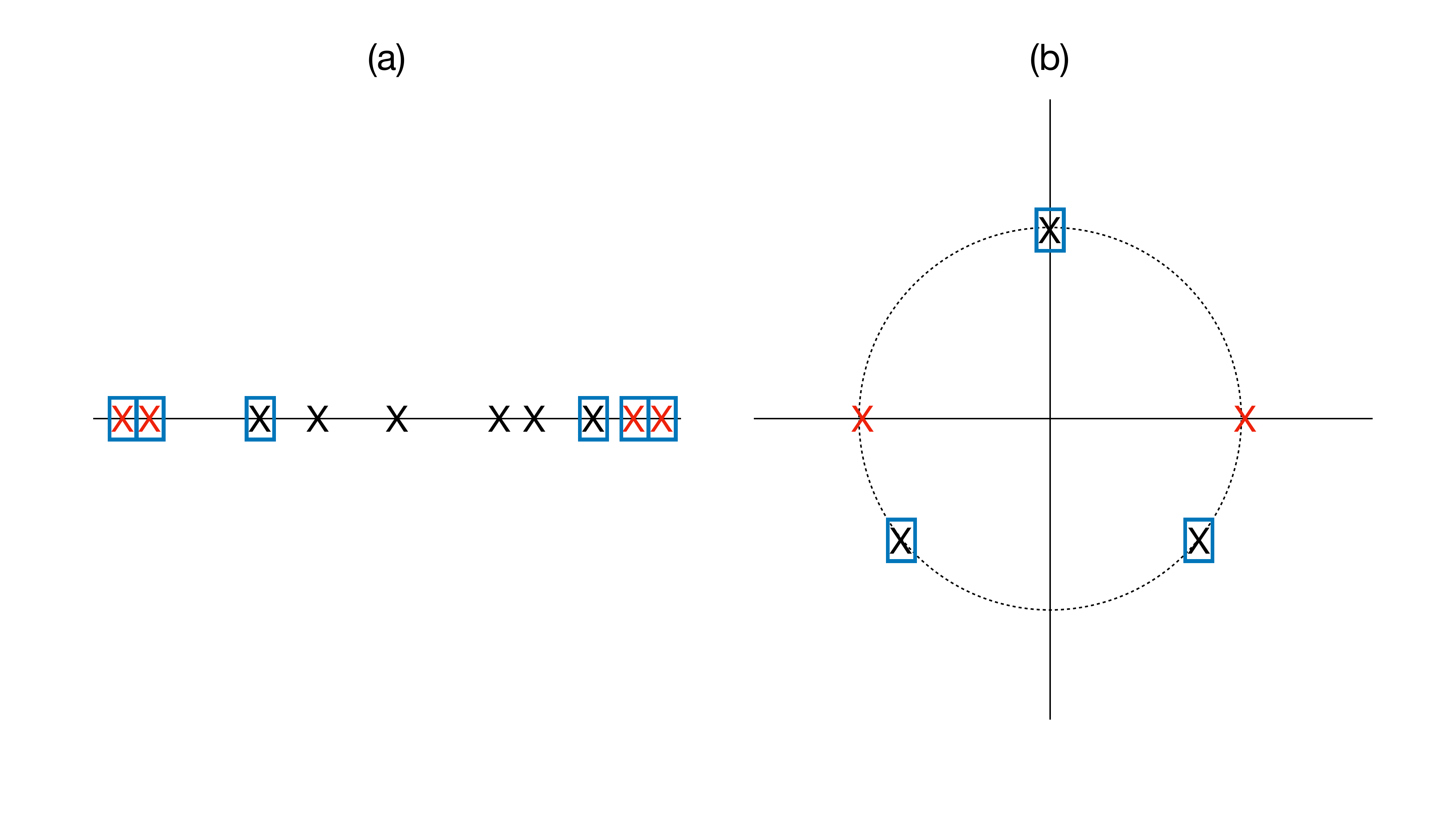}
    \caption{(a) $N=10$ random traits (`$\times$') on a line ($d=1$). In red, the subset that optimises the $M$-dispersion with $M=4$. The blue boxes indicate the subset that optimises the $M$-dispersion for $M=6$. The optimising subsets in $d=1$ are composed by two separate clusters for every $M$, comprising the $k$ leftmost and the $M-k$ rightmost variables for certain $k$'s. The optimal subset for $M+1$ includes the optimal subset for $M$, for any $M$. (b) $N=5$ random traits (`$\times$') on the plane ($d=2$). In red, the subset that optimises the $M$-dispersion with $M=2$. The blue boxes indicate the subset that optimises the $M$-dispersion for $M=3$. In dimension $d>1$, the optimal subset for $M+1$ does not necessarily include the optimal subset for $M$, as a global rearrangement could be more favorable in terms of increased dispersion.}
    \label{fig:schematics}
\end{figure}

Hence the maximiser grows recursively.  
Denote by
\begin{equation}
\bar x = \frac{1}{M}\sum_{i=1}^N x_i \sigma^{N,M}_i
\end{equation}
the mean of the optimal $M$-set $\sigma^{N,M}$.
The $(M+1)$-optimal set is obtained by adjoining the extremal point farthest from the current mean $\bar x$:
\begin{equation}
\sigma^{N,M+1} = \sigma^{N,M} + e_{i^*},\qquad 
i^*
=\argmax_{i\in\{k+1,M-k-1\}} |x_i-\bar x|\ .
\end{equation}
Thus, in $d=1$, optimal subsets are nested: the $(M+1)$-set contains the $M$-set. For $d>1$, however, this nesting property fails: adding one point may require a global rearrangement of the selected set (see Fig.~\ref{fig:schematics}).

In the following, we will first focus on the large-$N,M$ limit of the problem (with $M/N=\alpha$), developing a mean-field theory for the distribution of maximal dispersion that relies on the geometry of the optimal subset. Later, we will tackle the finite $N,M$ case, which requires more delicate considerations.

\subsection{Mean-field approach for large-$N,M$}\label{sec:meanfieldD1}

Let $X_1,\ldots,X_N$ be i.i.d. real random variables with common distribution function $F(x)$ whose density is $f(x)$. We denote by $ X_{(1)}, \ldots,X_{(N)} $ their order statistics, $X_{(1)} \leq X_{(2)} \leq \cdots \leq X_{(N)}$. Fix the fraction of leftmost and rightmost points  $p_L,p_R>0$ with $p_L+p_R=\alpha$, and define
\begin{equation}
\label{eq:index_set}
\mathcal I_N(p_L,p_R)=\{1,\ldots,\lfloor p_LN\rfloor\ \}\sqcup\{ N-\lceil p_RN\rceil+1,\ldots,N\}\ ,
\end{equation}
where $\lfloor\cdot\rfloor$ and $\lceil\cdot\rceil$ denote floor and ceiling functions, respectively.

For a symmetric kernel $h(x,y)$ vanishing on the diagonal, set
\begin{equation}\label{eq:S}
S=\sum_{i,j\in\mathcal I_N(p_L,p_R)}h(X_{(i)},X_{(j)})\ .
\end{equation}
In the limit $N\to\infty$, the empirical distribution of the i.i.d. random variables $X_1,\ldots,X_N$, converges to their common parent distribution $F$, and the rescaled sum $S/N^2$ in~\eqref{eq:S} converges to its mean-field limit
\begin{equation}\label{eq:limit_S}
\frac{1}{N^2}S\;\to\;
\iint_{((-\infty,a)\sqcup(b,\infty))^2} h(x,y)f(x)f(y)\,\de x\,\de y\ ,
\end{equation}
provided that the integral on the r.h.s. converges. The edge points $a,b$ are the quantiles of the distribution $F$ corresponding to the fractions $p_L$ and $p_R$; they are given by
\begin{equation}\label{eq:edgepoints}
F(a)=p_L ,\qquad 1-F(b)=p_R\ ,
\end{equation}
or, equivalently $a=G(p_L)$, $b=G(1-p_R)$, with $G=F^{-1}$ the functional inverse of $F$. 

Specialising equation~\eqref{eq:limit_S} to the maximal dispersion case corresponding to $h(x,y)=(x-y)^2$, we obtain the mean-field expression

\begin{equation}\label{eq:asymp_D1}
\frac{1}{N^2}D_{\max}^M\;\to\;\sup_{a,b:F(b)-F(a)=1-\alpha}\;
\iint\limits_{((-\infty,a)\sqcup(b,+\infty))^2}(x-y)^2 f(x)f(y)\,\de x\,\de y\ ,
\end{equation}
which is finite for sufficiently light-tailed $f$. 

The problem of maximising~\eqref{eq:asymp_D1} over all possible choices of $a,b$, reduces to the optimisation in a single variable $p_L$, with $p_R=\alpha-p_L$, as
\begin{equation}\label{eq:asymp_D}
\kappa_1^{(1)}(\alpha)=\sup_{0\leq p_L\leq\alpha}\;
\iint\limits_{\Omega(p_L;\alpha)}(x-y)^2 f(x)f(y)\,\de x\,\de y\ ,
\end{equation}
where
\begin{equation}
    \Omega(p_L;\alpha)=((-\infty,G(p_L))\sqcup(G(1-\alpha+p_L),+\infty))^2\ .
\end{equation}
The limit in~\eqref{eq:asymp_D} is equal to the (scaled) average value $\kappa_1^{(1)}(\alpha)$ defined in \eqref{eq:kappalimit}.

Differentiating the integral in \eqref{eq:asymp_D}, and using the identity $\frac{\de}{\de p_L}G(p_L)=\frac{1}{f(G(p_L))}$, yields
\begin{multline}\label{eq:derivative}
   2 \int_{-\infty}^{G(p_L)} (x-G(p_L))^2f(x)\,\de x
 - 2 \int_{G(1-\alpha+p_L)}^{+\infty} (x-G(1-\alpha+p_L))^2 f(x)\,\de x \\
 +2 \int_{G(1-\alpha+p_L)}^{+\infty} (x-G(p_L))^2 f(x)\,\de x
 -2 \int_{-\infty}^{G(p_L)} (x-G(1-\alpha+p_L))^2 f(x)\,\de x\ .
\end{multline}

The symmetry $f(x)=f(-x)$ implies we have $G(1-p)=-G(p)$. Therefore, the derivative vanishes at $p_L=\alpha/2$. In summary, 
\begin{equation}\label{eq:asymp_D_alpha}
\kappa_1^{(1)}(\alpha)=\iint\limits_{\Omega^\star}
(x-y)^2 f(x)f(y)\,\de x\,\de y\ ,
\end{equation}
where \begin{equation}\Omega^\star=\Omega(\alpha/2;\alpha) \ .\label{eq:Omegastar1d}
\end{equation}
This implies that in the mean-field limit, the optimal dispersion is realised by a \emph{balanced} configuration where the leftmost and rightmost blocks have equal weight. By the symmetry of $f$, formula~\eqref{eq:asymp_D_alpha} can be further simplified, and we have
\begin{equation}
\label{eq:Dmax:simpl_1d}
    \kappa^{(1)}_1(\alpha)=4\alpha\int_{-G(\alpha/2)}^{+\infty}x^2 f(x)\de x\ .
\end{equation}

\begin{ex}[Uniform distribution in an interval]
Suppose that the points $X_1,X_2,\ldots$ are uniformly distributed in the symmetric interval $[-1/2,1/2]$. In this case $F(x)=1/2+x$ for $x\in[-1/2,1/2]$ and $G(p)=p-\frac{1}{2}$. From~\eqref{eq:Dmax:simpl_1d} we get
\begin{equation} \label{eq:D_unif1d_cont}
    \kappa_1^{(1)}(\alpha) = 4 \alpha \int_{\frac{1-\alpha}{2}}^{1/2}x^2 \de x  = \frac{\alpha^2(\alpha^2-3\alpha+3)}{6}\ .
\end{equation}
\end{ex}

\subsection{Back to finite $N,M$}\label{sec:backfiniteN}
To compute the distribution of the maximal dispersion for finite $N,M$, one should evaluate 
\begin{equation}
\mathrm{Pr}\left(D_{\mathrm{max}}^M<x\right)=\mathrm{Pr}\left(D^M(1)<x,\ldots,D^M(M-1)<x\right) \ ,\label{eq:cumdistrMaxDexact}
\end{equation}
where with  $(D^M(1),\ldots,D^M(M-1))$ we denote the dispersions correspond to prefix-suffix configurations with $k=1,\ldots,M-1$ points in the leftmost interval. This evaluation is however technically challenging in general. We therefore resort to an approximate but very accurate evaluation which is asymptotically exact for large $N,M$.

Recall that, as $N \to \infty$, the prefix-suffix optimal subset contains \emph{balanced} fractions $p_L=p_R=\alpha/2$ on the left and on the right. For the dispersion of the balanced configuration, using the notation~\eqref{eq:index_set}, 
\begin{equation}
\label{eq:D_balanced}
D_{\mathrm{bal}}^M=\sum_{i,j\in{\mathcal{I}_N(\alpha/2,\alpha/2)}}\left(X_{(i)}-X_{(j)}\right)^2\ ,
\end{equation}
the following holds
\begin{equation}
    \lim_{\substack{N,M\to\infty\\ M/N=\alpha}}\left|\frac{1}{N^2}D_{\max}^M-\frac{1}{N^2}D_{\mathrm{bal}}^M\right|=0\ .
\end{equation}

Hence, the rescaled dispersion of a balanced subset is \emph{asympotically close} to the rescaled maximal dispersion. However, note that this property does not hold for the finite $N,M$ case: the configuration that realises the optimal dispersion is always a prefix-suffix, but not necessarily a balanced one.
\begin{ex}
    A small-$N$ example might help. Consider the problem of finding the maximal $M$-dispersion for the (deterministic) points  $x_1<x_2<\cdots<x_N$ with $x_{N-k+1}=2^{-(k-1)}$, for all $k=1,\ldots, N$. If $N=12$ and $M=8$, it is easy to check that the maximal dispersion is realized by the prefix-suffix choice 
    $$\{x_1,x_2,x_3,x_4,x_5,x_6\}\sqcup\{x_{11},x_{12}\}=\{1/2048, 1/1024, 1/512, 1/256, 1/128,1/64\}\sqcup\{1/2, 1\}\ ,$$
    with a dispersion of $32125735\times 2^{-21}$; this is larger than the dispersion of the balanced prefix-suffix selection 
    $$\{x_1,x_2,x_3,x_4\}\sqcup\{x_9,x_{10},x_{11},x_{12}\}=\{1/2048, 1/1024, 1/512,1/256\}\sqcup\{1/8,1/4, 1/2, 1\}$$
    corresponding to a dispersion of $29704135\times 2^{-21}$.
\end{ex}

\begin{figure}[h]
    \centering
    \includegraphics[width=.75\linewidth]{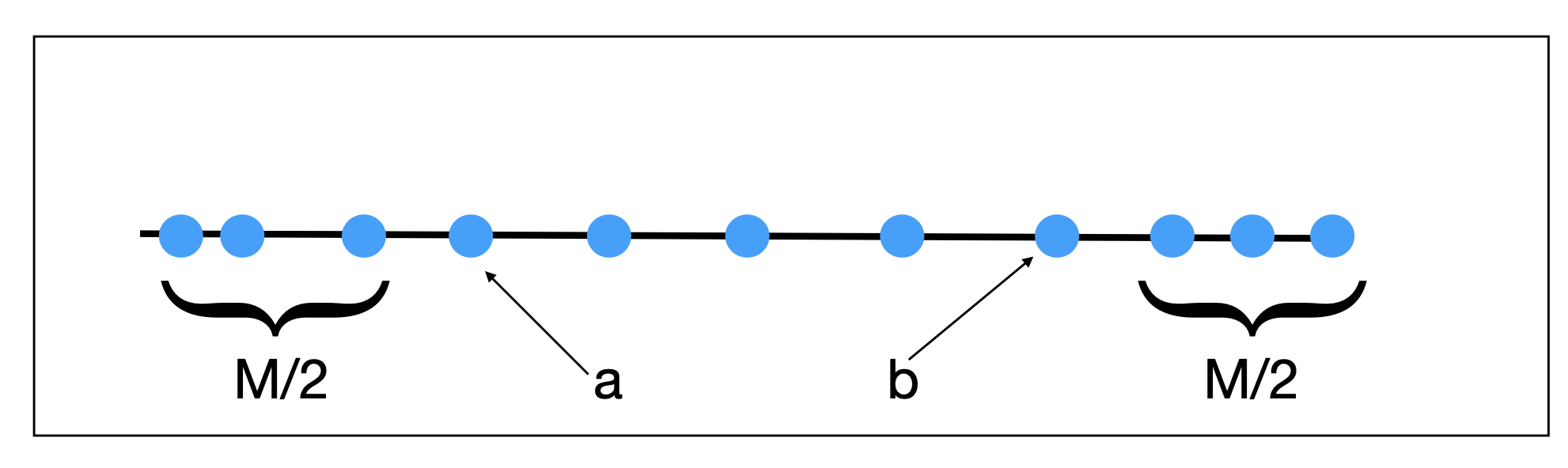}
    \caption{Schematic representation of the various terms of the expression in Eq. \eqref{eqmaintext:expgbalanced}.}
    \label{fig:1dOrder}
\end{figure}

Therefore, finite-$N$ approximants for $D_{\mathrm{max}}^M$ can be obtained from exact formulae for $D_{\mathrm{bal}}^M$, which are comparably easier to obtain than \eqref{eq:cumdistrMaxDexact}. To do that, we resort to the representation~\eqref{eq:D_balanced} in terms of the first $\lfloor M/2\rfloor$ and last $\lceil M/2\rceil$ order statistics. We summarise a few generalities on the distribution of the order statistics for i.i.d. random variables in~\ref{app:order_stat_generalities}. Using those formulae, we can write for a general function $g(D^M_{\mathrm{bal}})$,
    \begin{align}
 \nonumber      &\mathbb{E}  [g(D^M_{\mathrm{bal}})]
     =c_{N,M}\int \de a \de b F(a)^{M/2}f(a)\left[F(b)-F(a)\right]^{N-M-2}f(b)(1-F(b))^{M/2}\\ 
\nonumber        &\int \de x_1\cdots \de x_{\lfloor M/2\rfloor}  \de x_{N-\lceil M/2\rceil +1}\cdots \de x_{N} \frac{f(x_1)}{F(a)}\cdots \frac{f(x_{\lfloor M/2\rfloor})}{F(a)}\frac{f(x_{N-\lceil M/2\rceil+1})}{1-F(b)}\cdots \frac{f(x_N)}{1-F(b)}\\
    &\times  \prod_{i=1}^{M/2-1}\theta(x_{i+1}-x_i)\theta(b-a)\prod_{j=b}^{N-1}\theta(x_{j+1}-x_j) g\left(\sum_{i,j=\mathcal{I}_{N}(\alpha/2,\alpha/2)} (x_i-x_j)^2\right)\ ,\label{eqmaintext:expgbalanced}
\end{align}
with
\begin{equation}
    c_{N,M}:=\frac{N!}{(N-M-2)!\lfloor M/2\rfloor!\lceil M/2\rceil!}\ .
\end{equation}
The interpretation of the formula may be clearer by looking at Fig. \ref{fig:1dOrder}. The variables $a$ and $b$ represent the random position of the $M/2+1$-th and $N-(M/2)$-th variables. We want to have $M/2$ variables to the left of $a$, and $M/2$ to the right of $b$ (balanced configuration), an event that occur with the probability in the first line of Eq. \eqref{eqmaintext:expgbalanced}. The probability of the prefix-suffix configuration given $a$ and $b$ requires the conditional density of order statistics split in two blocks, which is given in Eq. \eqref{eq:1d_orderstat_conditioned} below, and inserted in the second line of Eq. \eqref{eqmaintext:expgbalanced}.  
Assuming $M$ even for notational simplicity, and taking $g(x)=x$, 
\begin{align}
\nonumber    \mathbb{E}[D_{\mathrm{bal}}^M]&=c_{N,M}\int_{-\infty}^{+\infty}  \de a\int_{a}^{+\infty}\de b F(a)^{M/2}f(a)\left[F(b)-F(a)\right]^{N-M-2}f(b)(1-F(b))^{M/2}\\
\nonumber   \Bigl\{ &\frac{M}{2}\left(\frac{M}{2}-1\right)\iint_{-\infty}^a \de x \de y \frac{f(x)}{F(a)}\frac{f(y)}{F(a)}(x-y)^2\\
\nonumber &+
2\left(\frac{M}{2}\right)^2\int_{-\infty}^a \de x\int_{b}^{+\infty}\de y \frac{f(x)}{F(a)}\frac{f(y)}{1-F(b)}(x-y)^2\\
&+
    \frac{M}{2}\left(\frac{M}{2}-1\right)\iint_{b}^{+\infty}\de x\de y \frac{f(x)}{1-F(b)}\frac{f(y)}{1-F(b)}(x-y)^2\Bigr\}\ .
\end{align}
Changing variables $F(x)=u$, $F(y)=v$ , and writing $n^{\underline{k}} := n (n-1) (n-2) \dots (n-k+1)$ for the falling factorial, 
\begin{align} 
\label{eq:1d_order_stat_Davg}
    \mathbb{E}[D^M_{\mathrm{bal}}] &=
    \nonumber c_{N,M}\int_{0}^{1}  \de a\int_{a}^{1}\de b~ a^{M/2}\left[b-a\right]^{N-M-2}(1-b)^{M/2}\\
   \nonumber &\times
   \Bigl\{ \left(\frac{M}{2}\right)^{\underline{2}}\iint_{-\infty}^a\frac{\de u}{a}\frac{\de v}{a} (G(u)-G(v))^2\\
   \nonumber &+
2\left(\frac{M}{2}\right)^2\int_{-\infty}^a \frac{\de u}{a}\int_{b}^{+\infty}\frac{\de v}{1-b} (G(u)-G(v))^2\\
&+
   \left(\frac{M}{2}\right)^{\underline{2}}\iint_{b}^{+\infty}\frac{\de u}{1-b}\frac{\de v}{1-b} (G(u)-G(v))^2\Bigr\} \ .
\end{align} 
A similar derivation provides the second moment,
\begin{align} 
\mathbb{E}[\left(D^M_{\mathrm{bal}}\right)^2]=c_{N,M}\int_{0}^{1}  \de a\int_{a}^{1}\de b~a^{M/2}\left[b-a\right]^{N-M-2}(1-b)^{M/2}H_M(a,b)\ ,\label{eq:1d_order_stat_2ndmomD}
\end{align}
where $H_M(a,b)$ is given by \eqref{eq:H_M} in \ref{appendix:H}. Subtracting the square of $\mathbb{E}[D^M_{\mathrm{bal}}]$ from \eqref{eq:1d_order_stat_2ndmomD} we get the variance. An alternative method to compute moments of $D_{\mathrm{bal}}^M$ is to use known moment formulae for order statistics~\cite{David54}.

\begin{ex}[Uniform distribution in $d=1$ at finite-$N$]
    For independent uniform random points in $[-1/2,1/2]$, we have $G(p)=p-1/2$. 
From \eqref{eq:1d_order_stat_Davg}, the finite-$N$ formula for the average dispersion of the balanced $M$-set reads
\begin{equation}
    \mathbb{E}\left[\frac{1}{N^2}D^{M}_{\mathrm{bal}} \right] =\frac{M \left(M^3-3 M^2 N-M^2+3 M N^2+3 M N-M-2\right)}{6 N^2(N+1) (N+2)}\ .
    \label{eq:DMax_mean_unif_finiteN}
\end{equation}
Setting $M=\alpha N$, we recover the value of the scaled first cumulant of the maximal $M$-dispersion computed in the mean-field approach~\eqref{eq:D_unif1d_cont},
\begin{equation}  
    \lim_{\substack{M,N \to \infty\\M/N=\alpha}} \mathbb{E}\left[\frac{1}{N^2}D^{M}_{\mathrm{bal}} \right]=\kappa_1^{(1)}(\alpha)  = \frac{\alpha^2(\alpha^2-3\alpha+3)}{6} \ .\label{largeNasymptKappa1Uniform}
\end{equation}
The comparison with the finite-$N$ formula is shown in Fig.~\ref{fig:finite_N_unif}. 
\begin{figure}
    \centering
    \includegraphics[width=\linewidth]{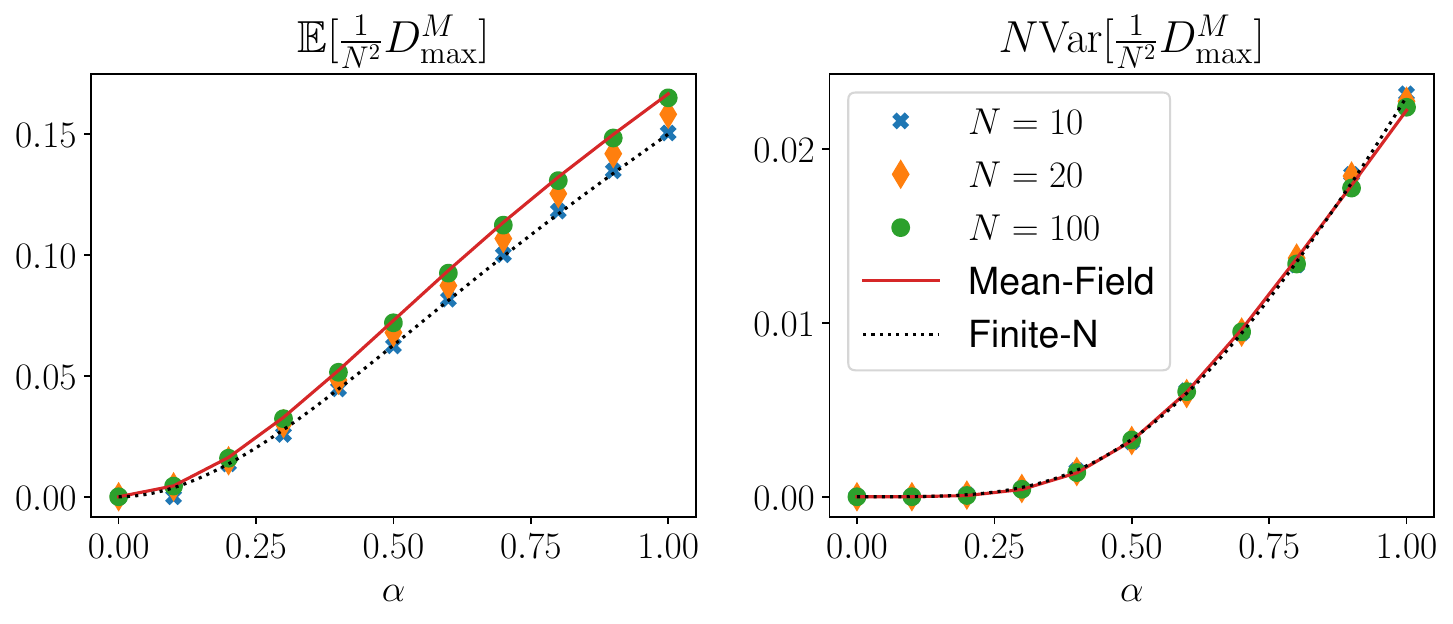}
    \caption{%
    Mean and variance of the maximal $M$-dispersion \textcolor{black}{for $d=1$, uniformly distributed random variables in $[-1/2,1/2]$} obtained numerically from the true (not necessarily balanced) prefix-suffix optimiser, compared with the theoretical predictions for the balanced configuration in the large $N$-limit provided by equations~\eqref{largeNasymptKappa1Uniform} and \eqref{eq:var_divmax_unif} (red line). The dotted line shows the predictions for the balanced configuration for the smallest value $N=10$, provided by equations~\eqref{eq:DMax_mean_unif_finiteN} and~\eqref{eq:DMax_variance_unif_finiteN} with $M=\lfloor \alpha N\rfloor$, which confirms that the balanced configuration is close to the optimal one already at such low value of $N$. The sample size for the estimation of mean and variance used in the numerics here is $10^4$.}
    \label{fig:finite_N_unif}
\end{figure} 
\par

As for the variance, from \eqref{eq:1d_order_stat_Davg} and \eqref{eq:1d_order_stat_2ndmomD}  we have
\begin{multline}
\operatorname{Var}\left[\frac{1}{N^2}D^{M}_{\mathrm{bal}}\right] =\frac{M (M+2)}{90 N^4(N+1)^2 (N+2)^2 (N+3) (N+4)}\times\\
 \bigl(-10 M^6 N-25 M^6+42 M^5 N^2+136 M^5 N+64 M^5-60 M^4 N^3-223 M^4 N^2-124 M^4 N\\
 +69 M^4+30 M^3 N^4+105 M^3 N^3-52 M^3 N^2-341 M^3 N-154 M^3+30 M^2 N^4 +210 M^2 N^3\\
 +454 M^2 N^2+272 M^2 N-17 M^2-20 M N^2-80 M N-90 M-36 N^2-108 N-72\bigr)\ .
 \label{eq:DMax_variance_unif_finiteN}
\end{multline}
Setting  $M=\alpha N$, and rescaling, we obtain the expression for the scaled second cumulant
\begin{equation}\label{eq:var_divmax_unif}
  \kappa_2^{(1)}(\alpha) = \lim_{\substack{M,N \to \infty\\M/N=\alpha}}N\operatorname{Var}\left[\frac{1}{N^2}D^{M}_{\mathrm{bal}}\right] =  \frac{30 \alpha^5 - 60 \alpha^6 + 42 \alpha^7 - 10 \alpha^8}{90}\ .
\end{equation}
See Fig.~\ref{fig:finite_N_unif} for numerical checks of these formulae. 
\end{ex}

\section{Any number of traits ($d \geq 1$) -- Mean field approach} \label{sec:continuous}

In  dimension $d=1$, the optimal $M$-subset can be described explicitly: for large $N$, it is, with high probability, the complement of an interval centered at the mean.  
This motivates the mean field formulation for $d \geq 1$, in the limit $N,M \to \infty$ with fixed $\alpha = M/N$ that will lead us to general formulae for (i) the typical value of $D_{\mathrm{max}}^M(\bm X_1,\ldots,\bm X_N)$ (subsection~\ref{subsec:typical}), and (ii) the Scaled Cumulant Generating Function and the rate function governing large fluctuations of the maximal dispersion (subsection~\ref{subsec:SCGF}).

\subsection{Typical maximal $M$-dispersion}
\label{subsec:typical}
By the very same considerations made for the $d=1$ case,  for general $d>1$, we expect the rescaled maximal $M$-dispersion to converge to its continuous (mean-field) analogue,
\begin{equation}
   \lim_{N \to \infty}\frac{1}{N^2}D_{\mathrm{max}}^M=\sup_{\Omega\subset\mathbb{R}^d\colon \int_{\Omega}f(\bm x)\de\bm x=\alpha}\iint_{\Omega\times\Omega}|\bm x-\bm y|^2 f(\bm x)f(\bm y)\de \bm x\de \bm y\ ,\label{continuousddim}
\end{equation}
provided that the integral on the r.h.s. converges. 
Equation~\eqref{continuousddim} is the analogue of~\eqref{eq:asymp_D} for $d>1$.

For $d=1$, the supremum is attained for
\begin{equation}
\Omega^* = (-\infty, G(\alpha/2)] \cup [G(1-\alpha/2), +\infty)\ ,
\end{equation}
(see \eqref{eq:Omegastar1d}), where $G = F^{-1}$ is the quantile function of the parent distribution $F$. What plays the role of $\Omega^*$ in higher dimensions?

We first rewrite the functional:
\begin{align}
\nonumber
\iint_{\Omega\times\Omega} |\bm x - \bm y|^2 f(\bm x) f(\bm y)\, \de\bm x\, \de\bm y
 &= 2\alpha \int_{\Omega} |\bm x|^2 f(\bm x)\, \de\bm x
   - 2 \textcolor{black}{\sum_{\ell=1}^d}\Big(  \int_{\Omega} x_\ell f(\bm x)\, \de\bm x \Big)^2 \\
 &= 2\alpha \int_{\Omega}
     \Big| \bm x - \frac{1}{\alpha} \int_{\Omega} \bm y f(\bm y)\, \de\bm y \Big|^2
     f(\bm x)\, \de\bm x\ .\label{eq:functionalOmega}
\end{align}
Thus the problem reduces to:  
given a density $f(\bm x)$ of unit mass, find the region $\Omega$ of measure $\alpha$ that maximises the moment of inertia of $f(\bm x)\mathds{1}_{\Omega}$ about its own center of mass.
\textcolor{black}{We conjecture—guided by the one-dimensional result and by the structure of the
functional \eqref{eq:functionalOmega} —that the maximal dispersion is asymptotically close to the dispersion realized on $\Omega^*$, the complement of a ball centered at:}
\begin{equation}
\mathbb{E} \left[\bm X\right]=\int_{\mathbb{R}^d} \bm x f(\bm x)\, \de\bm x\ ,
\end{equation}
with radius $R(\alpha)>0$ determined by
\begin{equation}
1-\alpha = \int_{|\bm x| \le R(\alpha)} f(\bm x)\, \de\bm x \ .
\end{equation}
Hence,
\begin{equation}
\Omega^* = \{ \bm x \in \mathbb{R}^d : |\bm x - \mathbb{E}[\bm X]| \ge R(\alpha) \}\ .
\end{equation}
If $f$ is rotationally symmetric, then $\mathbb{E} [\bm X]=0$, and
\begin{equation}
   \label{eq:general_continuous_formula}
   \kappa_1^{(d)}(\alpha):=   \lim_{N \to \infty}\frac{1}{N^2}\mathbb{E}[ D_{\mathrm{max}}^M]
   = 2\alpha \int\limits_{|\bm x| \ge R(\alpha)} |\bm x|^2 f(\bm x)\, \de\bm x\ ,
\end{equation}
where
\begin{equation}\label{eq:R_alphaSelfCons}
   \int\limits_{|\bm x| \ge R(\alpha)} f(\bm x)\, \de\bm x = \alpha\ .
\end{equation}

\begin{figure}
    \centering
    \includegraphics[width=0.99\linewidth]{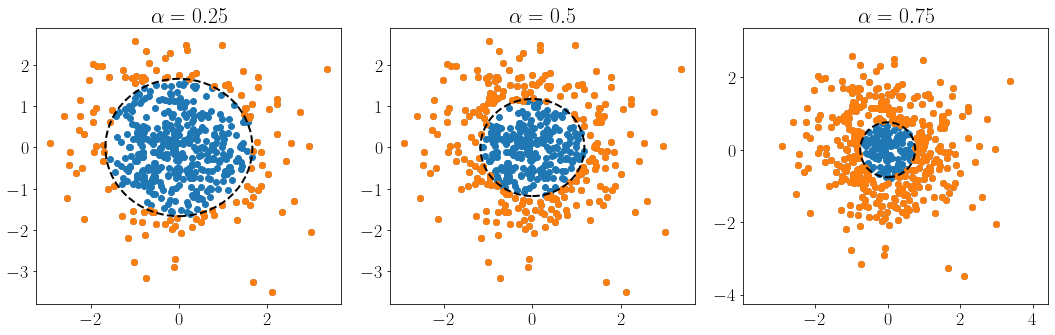}
    \caption{Plots for $N=500$ points $(x,y)$ sampled according to the standard Gaussian distribution in $d=2$. The yellow points are producing the maximal $M$-dispersion for different values of $\alpha=M/N$ according to the greedy algorithm described in Sec.~\ref{sec:heuristics}. The black dashed circle centered in the origin has radius $R(\alpha)$ given by \eqref{eq:RadiusFabio}.}
    \label{fig:selectedpoints2dGaussian}
\end{figure}

\begin{ex}[Gaussian distributions in $d>1$]
  For the standard Gaussian density
\begin{equation}
 f(\bm x) = \frac{1}{(2\pi)^{d/2}} \e^{-|\bm x|^2/2},  \quad \bm x\in\R^d\ , 
\end{equation}
formula~\eqref{eq:general_continuous_formula} becomes
\begin{equation}
  \kappa_1^{(d)}(\alpha)= 2\alpha\int\limits_{|\bm x|\geq R(\alpha)}|\bm x|^2 f(\bm x)\de\bm x= 2\alpha\frac{2^{1-\frac{d}{2}}}{\Gamma \left(\frac{d}{2}\right)}\int\limits_{R(\alpha)}^{+\infty}r^{d+1}\e^{-r^2/2}\de r=
    \frac{4\alpha}{\Gamma \left(\frac{d}{2}\right)} \Gamma \left(\frac{d}{2}+1,\frac{R(\alpha)^2}{2}\right), 
   \label{eq:D_vs_alpha_Gauss}
\end{equation}
where $R(\alpha)$ is the solution of the equation~\eqref{eq:R_alphaSelfCons}
\begin{equation}
   \Gamma\!\left(\dfrac{d}{2}, \dfrac{R(\alpha)^2}{2}\right)
   = \alpha\, \Gamma\!\left(\dfrac{d}{2}\right)\ .
   \label{eq:R_vs_alpha_Gauss}
\end{equation}
In the two-dimensional case  ($d=2$), this reduces to
\begin{equation}\label{eq:RadiusFabio}
    \e^{-R(\alpha)^2/2}=\alpha \Rightarrow R(\alpha)=\sqrt{2\log(1/\alpha)}\ .
\end{equation}
Hence
\begin{equation}
    \kappa_1^{(2)}(\alpha)
    =4 \alpha^2  \left(1-\log\alpha\right)\ .\label{kappa12GaussFromAverageMeanField}
\end{equation}
\end{ex}
We used a heuristic numerical algorithm described in Sec.~\ref{sec:heuristics} to identify, for Gaussian samples in $d=2$, the $M$-subset that maximises the dispersion for various values of $\alpha$. 
The resulting configurations, shown in Fig.~\ref{fig:selectedpoints2dGaussian}, coincide with the theoretical prediction: 
the boundary of the optimal region $\Omega^*$ aligns with the circle of radius $R(\alpha)$ given by~\eqref{eq:RadiusFabio}.

\begin{figure}
    \centering    \includegraphics[width=0.99\linewidth]{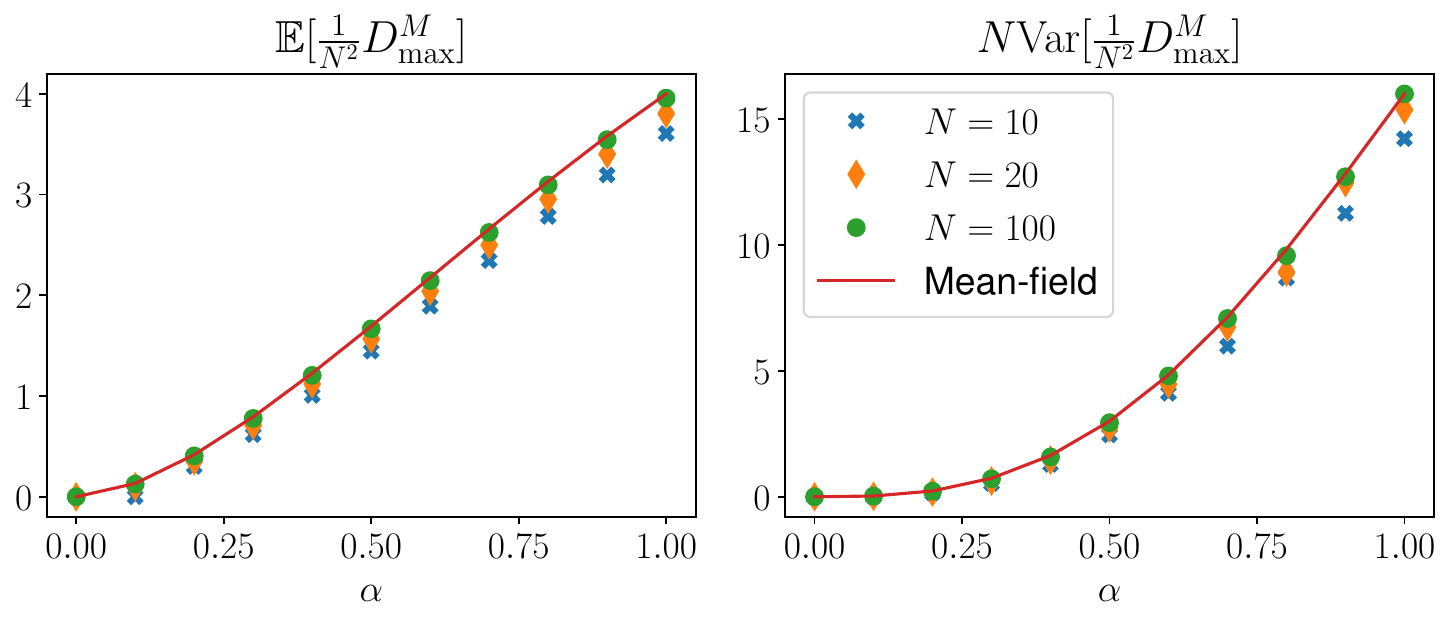}
    \caption{Mean and variance of $D_{\mathrm{max}}^M$ (with appropriate normalisation in $N$) for Gaussian points in $d=2$. The results obtained from the greedy algorithm with several values of $N$ are compared with the theoretical values (red line). The numerical estimates have been obtained from a sample of size  $10^4$.}
    \label{fig:enter-label}
\end{figure}

\subsection{SCGF and Rate Function of the maximal $M$-dispersion}
\label{subsec:SCGF}
The analysis for large $N$, which revealed the typical value of $D_{\mathrm{max}}^M$, can be extended to investigate its fluctuations. One may think of the typical configuration as the most probable spatial arrangement of points: a fraction $\alpha$ of them lie beyond a certain radius $R(\alpha)$, itself determined by the self-consistent relation
\begin{equation}
\int_{|\bm{x}|\geq R(\alpha)} f(\bm{x})\,\de\bm{x}=\alpha\ .
\end{equation}
The corresponding limiting form of $D_{\mathrm{max}}^M$ follows:
\begin{align}
\label{eq:typical_D_conf}
    &\frac{1}{N^2}D^M_{\max}\simeq\frac{1}{N^2}\sum_{i,j\colon |\bm{X}_i|,|\bm{X}_j|\geq R(\alpha)}|\bm{X}_i-\bm{X}_j|^2\\
    &=2\frac{M}{N^2}\sum_{i\colon |\bm{X}_i|\geq R(\alpha)}\left|\bm{X}_i\right|^2-2\left|\frac{1}{N}\sum_{i\colon |\bm{X}_i|\geq R(\alpha)}\bm{X}_i\right|^2
    \to 2\alpha\int_{|\bm{x}|\geq R(\alpha)} |\bm x|^2f(\bm{x})\,\de \bm{x}\ ,
\end{align}
since by rotational symmetry, the second term vanishes.  

To go beyond the typical value, consider the Laplace transform
\begin{align}
    \mathbb{E}[\e^{-p D^M_{\max}/N}] =
    \int_{\R^d}\!\!\cdots\!\!\int_{\R^d} 
    f(\bm{x}_1)\cdots f(\bm{x}_N) 
    \exp\!\left[-\frac{p}{N}D_{\mathrm{max}}^M\right]
    \,\de\bm{x}_1\cdots \de\bm{x}_N\ .
    \label{eq:Laplace}
\end{align}
For large $N$, this integral is dominated by those configurations that realise the most likely dispersion.  
At $p=0$, we recover the typical geometry described earlier, in which a fraction $\alpha$ of the points lies outside the ball of radius $R(\alpha)$.

When $p\neq0$, the picture tilts.  
For $p>0$, the weight favors configurations with atypically small $D_{\mathrm{max}}^M$; for $p<0$, it favors those with atypically large dispersion.  
We assume, by analogy with the typical case, that these atypical configurations are again radially organised, characterised by a fraction $\alpha$ of points lying beyond some radius $R(\alpha,p)$, smaller or larger than $R(\alpha)$ according to the sign of $p$.

In such a configuration, one expects
\begin{align}
    &\frac{1}{N}\sum_{i,j\colon |\bm{X}_i|,|\bm{X}_j|\geq R(\alpha,p)}|\bm{X}_i-\bm{X}_j|^2
    \sim 2\alpha \sum_{i\colon |\bm{X}_i|\geq R(\alpha,p)}\left|\bm{X}_i\right|^2\ ,
\end{align}
so that to each $p$ there corresponds an optimal radius $R(\alpha,p)$ balancing the competing weights.  Thus, asymptotically,
\begin{align}
  \mathbb{E}[\e^{-p D^M_{\max}/N}]
  &\approx
 {\sup_{R\geq0}} 
 \binom{N}{M}
 \left(\int_{|\bm x|\leq R}f(\bm x)\,\de\bm x\right)^{N-M}
 \left(\int_{|\bm x|\geq R}f(\bm x)\e^{-2\alpha p |\bm x|^2}\,\de\bm x\right)^{M}\ .
    \label{eq:gen_ratef}
\end{align}

In the joint limit $N,M\to\infty$ with $M/N\to\alpha$, the combinatorial term contributes the entropy
\begin{equation}
   \lim_{N\to\infty}-\frac{1}{N} \log\binom{N}{M}
   =\alpha\log\alpha+(1-\alpha)\log(1-\alpha)\ .
\end{equation}
Consequently, the transform \eqref{eq:gen_ratef} assumes the large-deviation form
\begin{equation}
     \mathbb{E}[\e^{-p D^M_{\max}/N}]
     \approx \e^{-N \inf_{R\geq0}L_{\alpha}(p,R)}\ ,
\end{equation}
with
\begin{align}
    L_\alpha(p,R)=
   -\alpha\log\left(\frac{1}{\alpha}\int_{|\bm{x}|\geq R}f(\bm{x})\e^{-2\alpha p |\bm{x}|^2}\,\de \bm{x}\right)
   -(1- \alpha)\log\left(\frac{1}{1-\alpha}\int_{|\bm{x}|\leq R}f(\bm{x})\,\de \bm{x}\right) \ .
   \label{eq:aux_SCGF_radial}
\end{align}
In this way, the SCGF emerges as the minimal value:
\begin{equation}
\label{eq:SCGF_radial}
  \Phi_\alpha(p)=\inf_{R\geq0}L_{\alpha}(p,R)=L_{\alpha}(p,R(\alpha,p))\ ,
\end{equation}
where the optimal radius $R(\alpha,p)$ satisfies the stationarity condition
\begin{equation}
     \frac{\partial }{\partial R}L_\alpha(p,R)\Big|_{R=R(\alpha,p)}=0\ ,
    \label{conditionRpDerivativeL}
\end{equation}
which reads explicitly as
\begin{equation}
\label{eq:R_sol}
    (1-\alpha)\frac{ 1}{\int_{|\bm{x}|\leq R(\alpha,p)}f(\bm{x})\,\de \bm{x}}
    -\alpha 
    \frac{\e^{-2\alpha p R(\alpha,p)^2}}{\int_{|\bm{x}|\geq R(\alpha,p)}f(\bm{x})\e^{-2\alpha p |\bm{x}|^2}\,\de \bm{x}}
    =0\ .
\end{equation}
At $p=0$, one recognises the equilibrium condition
\begin{equation}
    \int_{|\bm{x}|\geq R(\alpha,0)}f(\bm{x})\,\de \bm{x}=\alpha\ ,
    \label{SCGFconditionforRzero}
\end{equation}
so that $R(\alpha,0)=R(\alpha)$.   
The structure is clear:
the Laplace principle selects a critical radius $R(\alpha,p)$ balancing two complementary masses of $f$, the outer one being  
 weighted by the exponential tilt.  
At $p=0$, this balance collapses to the simple geometric $\alpha$-quantile condition $ R(\alpha,0)=R(\alpha)$ defining the typical state.

With \eqref{eq:R_sol}, the SCGF takes the compact form
\begin{equation}
    \Phi_\alpha(p)
    =\log (1-\alpha )
    +2 \alpha ^2 p R(\alpha,p)^2
    -\log \left(\int_{|\bm{x}|\leq R(\alpha,p)}f(\bm{x})\,\de \bm{x}\right)\ .
    \label{SimplifiedPhiAlphapSCGF}
\end{equation}
Derivatives of $\Phi_\alpha(p)$ at $p=0$ give the leading cumulants (see Eq. \eqref{eq:leadingcumulantsfromPhi}).  
Since $\partial_R L_\alpha=0$ along the optimal path, we have
\begin{align}
\label{eq:1st_derivative}
    \kappa_1^{(d)}(\alpha)
    =\frac{\de}{\de p}\Phi_{\alpha}(0)
    =\frac{\partial L_\alpha}{\partial p}
    +\underbrace{\frac{\partial L_\alpha}{\partial R}}_{=0}\frac{\de R}{\de p}
    =\frac{\partial L_\alpha}{\partial p}\biggl|_{(0,R(\alpha,0))}\ ,
\end{align}
while, to leading order, the variance reads
\begin{align}
\label{eq:2nd_derivative}
\kappa_2^{(d)}(\alpha)
=-\frac{\de^2}{\de p^2}\Phi_{\alpha}(0)
=-\frac{\partial^2 L_\alpha}{\partial p^2}\biggl|_{(0,R(\alpha,0))}
+\frac{\left(\frac{\partial^2 L_\alpha}{\partial R\partial p}\right)^2}{\frac{\partial^2 L_\alpha}{\partial R^2}}\biggl|_{(0,R(\alpha,0))}\ .
\end{align}

\begin{ex}[Uniform distribution in $d=1$]
   Consider the case of uniform density in the interval $[-1/2,1/2]$.
     For $0\leq R\leq 1/2$,
 \begin{equation}
 \int_{|x|\leq R}\mathds{1}_{[-1/2,1/2]}(x)\de x=2R\ ,
\end{equation}
and
\begin{equation}
 \int_{|x|\geq R}\mathds{1}_{[-1/2,1/2]}(x)\e^{-2\alpha p |x|^2}\de x=\sqrt{\frac{\pi}{2}}\frac{1}{\sqrt{\alpha p}}\left(\operatorname{erf}\left(\sqrt{\frac{\alpha p}{2}}\right)-\operatorname{erf}\left(\sqrt{{2\alpha p}}R\right)\right)\ .\label{unif1dSCGFExpp}
  \end{equation}

From \eqref{SimplifiedPhiAlphapSCGF}, the SCGF is
  \begin{equation}
      \Phi_\alpha(p) = \log(1-\alpha)+2\alpha^2 p R(\alpha,p)^2
-\log(2R(\alpha,p))\ ,\label{eq:SCGFuniform1d}
\end{equation}
 where $R(\alpha,p)$ is the solution of
 \begin{equation}
 \label{eq:R_unif}
   \alpha\frac{2\e^{-2\alpha p R(\alpha,p)^2}\sqrt{2\alpha p/\pi}}{\operatorname{erf}\left(\sqrt{\frac{\alpha p}{2}}\right)-\operatorname{erf}\left(\sqrt{{2\alpha p}}R(\alpha,p)\right)}  -\frac{1-\alpha}{R(\alpha,p)}=0\ .
 \end{equation} 
Using the limit
\begin{equation}
   \lim_{p\to 0}\frac{\sqrt{\frac{2 \alpha  p}{\pi }}}{\mathrm{erf}\left(\sqrt{\frac{\alpha  p}{2}}\right)-\mathrm{erf}\left(R \sqrt{2 \alpha  p}\right)}=\frac{1}{1-2 R}\ ,
\end{equation}
we deduce that $R(\alpha)=(1-\alpha)/2$, which immediately implies $\Phi_\alpha(0)=0$ as expected by normalisation. It can be verified that $\Phi_{\alpha}'(0)$ and $-\Phi_{\alpha}''(0)$ coincide with the leading order of the average value in~\eqref{largeNasymptKappa1Uniform} and the variance~\eqref{eq:var_divmax_unif} respectively.
\begin{figure}[t]
    \centering
\includegraphics[width=0.485\linewidth]{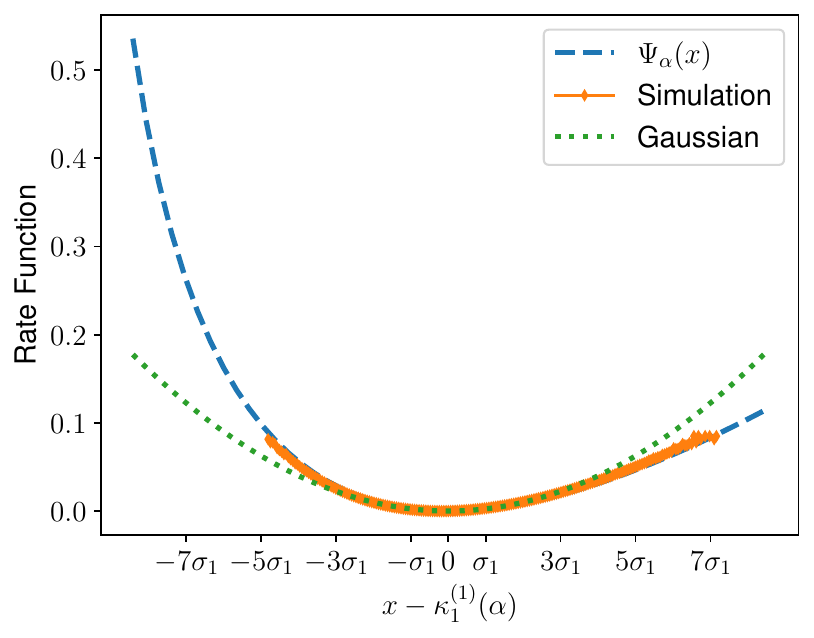}~
\includegraphics[width=0.49\linewidth]{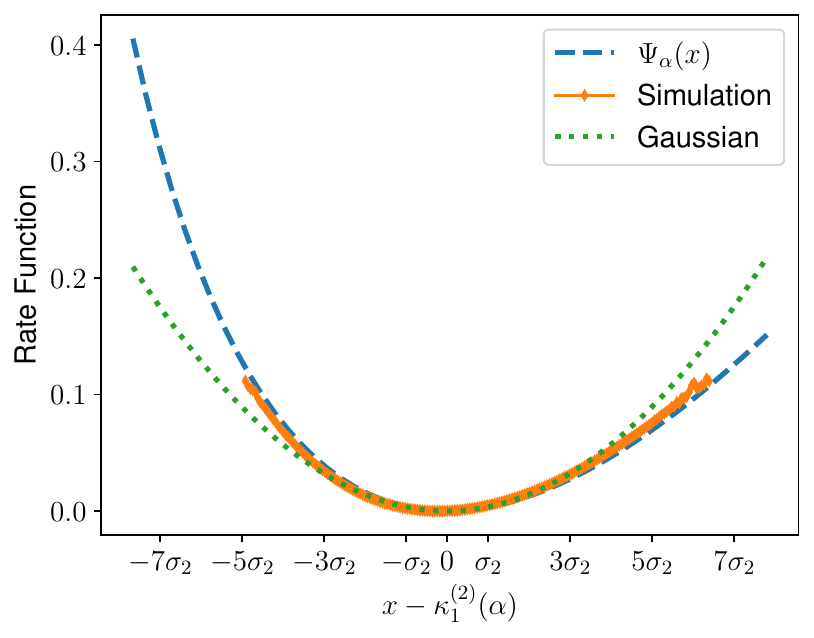}
    \caption{Comparison between analytical results and a numerical estimation of the large deviation tails. The numerical results in both cases are obtained with $10^9$ samples. (Left) Results obtained for the Gaussian case in $d=1$ with $N=200$ and $M=80$. The blue dashed line is obtained from a numerical evaluation of the Legendre-Fenchel transform of \eqref{eq:SGF_1d_gauss_cont}, while the yellow solid line is obtained from a numerical simulation using an exact algorithm for the evaluation of $D^M_{\mathrm{max}}$. The green dotted line shows the rate function for a Gaussian variable with average and variance given respectively by \eqref{eq:aveScaledGauss1d} and \eqref{eq:varScaledGauss1d}. (Right) Results obtained for the Gaussian case in $d=2$ with $N=140$ and $M=80$. The blue dashed line is obtained from a numerical evaluation of the Legendre-Fenchel transform of \eqref{eq:SGF_2d_gauss_cont}, while the yellow solid line is obtained from a numerical simulation using a greedy algorithm for the evaluation of $D^M_{\mathrm{max}}$. The green dotted line shows the quadratic behaviour for a Gaussian variable with average and variance given respectively by \eqref{eq:aveScaledGauss2d} and \eqref{eq:varScaledGauss2d}.}
    \label{fig:rateFunction}
\end{figure}
\end{ex}

\begin{ex}[Gaussian density]
    Consider again the $d$-dimensional standard Gaussian density
\begin{equation}
    f(\bm x)=\frac{1}{\left(2\pi\right)^{d/2}}\e^{-|\bm x|^2/2}\ ,\quad \bm{x}\in\R^d\ .
\end{equation}
Then,
\begin{align}
  \int_{|\bm{x}|\leq R}f(\bm{x})\de \bm{x}&= 1- \frac{\Gamma \left(\frac{d}{2},\frac{R^2}{2}\right)}{\Gamma \left(\frac{d}{2}\right)}\\
   \int_{|\bm{x}|\geq R}f(\bm{x})\e^{-2\alpha p|\bm{x}|^2}\de \bm{x}&= \frac{2^{-\frac{d}{2}}}{\Gamma\left(\frac{d}{2}\right)} R^d E_{1-\frac{d}{2}}\left(\frac{1}{2} (1+4\alpha p) R^2\right) ,
\end{align}
where $E_n(z):=\int_{1}^{\infty}\e^{-zt}t^{-n}\de t$ is the exponential integral function. 
Inserting those integrals in~\eqref{eq:R_sol}-\eqref{SimplifiedPhiAlphapSCGF}, we get the SCGF. In the following, we specialise these formulae to the $d=2$ and $d=1$ cases.

\begin{enumerate}
\item {\bf Dimension $d=2$}\\ 
\\
  The solution of Eq.~\eqref{eq:R_sol} is 
  \begin{equation}
 R(\alpha,p)=\sqrt{2\log\left(\frac{1+4 \alpha ^2 p}{  \alpha+4 \alpha^2  p}\right)}\ ,
  \end{equation}
  with the SCGF from~\eqref{eq:SCGF_radial} given by 
 \begin{align}
     \Phi_{\alpha}(p)=\log (1-\alpha )+4 \alpha ^2 p \log \left(\frac{1+4 \alpha ^2 p}{\alpha +4 \alpha ^2 p}\right)-\log \left(\frac{1-\alpha}{1+4 \alpha ^2 p}\right)\ .\label{eq:SGF_2d_gauss_cont}
 \end{align}
 
By normalisation, $\Phi_\alpha(0)=0$, and the condition  $p>-1/(4\alpha)$ is required for the function to be real-valued. Taking  derivatives w.r.t $p$ and setting $p=0$, we obtain the leading order of the cumulants:  
 \begin{align}\label{eq:aveScaledGauss2d}
     \kappa_1^{(2)}(\alpha) &= 4\alpha^2(1-\log\alpha)\\
     \label{eq:varScaledGauss2d}
     \kappa_2^{(2)}(\alpha) &= 16\alpha^3 (2-\alpha)\\
         \kappa_{\ell}^{(2)}(\alpha) &= 4^{\ell}(\ell-2)!\alpha^{\ell+1} (\ell-\alpha^{\ell-1}),\quad \ell\geq 3\ .
 \end{align}
 While the leading term of the average was already obtained in Eq. \eqref{kappa12GaussFromAverageMeanField}, the leading order of the variance and higher cumulants are new results. 

The rate function in real space $\Psi_\alpha (x)=\sup_p \left(\Phi_\alpha(p)-px\right)$  is plotted in Fig.~\ref{fig:rateFunction} along with results of numerical simulations. The deviations from the typical fluctuations (Gaussian) around the mean are evident in the tails.

\item {\bf Dimension $d=1$}\\ 
\\
  For the standard Gaussian density in $d=1$
\begin{equation}
 f(x) = \frac{1}{(2\pi)^{1/2}} \e^{-x^2/2}\ ,
\end{equation}
these formulae specialise to
 \begin{equation}
    \int_{|\bm{x}|\leq R}f(\bm{x})\de \bm{x}\equiv \int_{-R}^R\de x~\frac{1}{(2\pi)^{1/2}} \e^{-x^2/2}=\operatorname{erf}\left(\frac{R}{\sqrt{2}}\right)\ ,
\end{equation}
and
\begin{equation}
 \int_{|\bm{x}|\geq R}f(\bm{x})\e^{-2\alpha p|\bm{x}|^2}\de \bm{x}=
 \int_{x\in \Omega(R)}\de x~\frac{1}{(2\pi)^{1/2}} \e^{-x^2/2-2\alpha p x^2}=\frac{\operatorname{erfc}\left(\frac{R}{\sqrt{2}}\sqrt{1+4\alpha p}\right)}{\sqrt{1+4\alpha p}}\ ,
  \end{equation}
in terms of the error function $\mathrm{erf}(z)=(2/\sqrt{\pi})\int_0^z\de t~\exp(-t^2)$ and complementary error function $\mathrm{erfc}(z)=1-\mathrm{erf}(z)$. Here, $\Omega(R)=(-\infty,-R)\sqcup (R,\infty)$.

Hence, the SCGF is
   \begin{align}
\nonumber     \Phi_{\alpha}(p)&=- \alpha\log\left(\frac{1}{\alpha}\frac{\operatorname{erfc}\left(\frac{R(\alpha,p)}{\sqrt{2}}\sqrt{1+4\alpha p}\right)}{\sqrt{1+4\alpha p}}\right)
   -(1- \alpha)\log\left(\frac{1}{1-\alpha}\operatorname{erf}\left(\frac{R(\alpha,p)}{\sqrt{2}}\right) \right)\\
&=2 \alpha ^2 p R(\alpha,p)^2-\log \left(\frac{\text{erf}\left(\frac{R(\alpha,p)}{\sqrt{2}}\right)}{1-\alpha }\right)\ ,\label{eq:SGF_1d_gauss_cont}
\end{align}
  where $R(\alpha,p)$ is determined by the condition \eqref{eq:R_sol}
  \begin{equation}
   (1-\alpha)\frac{ 1}{\operatorname{erf}\left(\frac{R(\alpha,p)}{\sqrt{2}}\right)}-\alpha \frac{\e^{-2\alpha p R(\alpha,p)^2}}{\frac{\operatorname{erfc}\left(\frac{R(\alpha,p)}{\sqrt{2}}\sqrt{1+4\alpha p}\right)}{\sqrt{1+4\alpha p}}}=0\ .\label{conditionfGauss1dSCGF} 
   \end{equation}
   In this case
   \begin{equation} \label{eq:R0_1d_gauss_continuous_SGF}
       R(\alpha,0)\equiv R(0)=\sqrt{2} \operatorname{erf}^{-1}(1-\alpha )
   \end{equation}
   in terms of the functional inverse of the error function, which leads to $\Phi_\alpha(0)=0$, by normalisation. Note also that the SCGF is defined again only for $p>-1/(4\alpha)$, which will lead in both cases ($d=1,2$) to a rate function $\psi_\alpha(x)$ that is defined only for $x>0$ as expected. 
Taking  derivatives w.r.t $p$ at $p=0$, we obtain the leading order of the cumulants. Since the expressions are rather lengthy,  we report only the first two:
\begin{equation}
    \kappa_1^{(1)}(\alpha)=\Phi_\alpha '(0)=2 \alpha ^2+\frac{4 \alpha~\e^{-(\tau(\alpha))^2} \tau(\alpha)}{\sqrt{\pi }}\ ,
    \label{eq:aveScaledGauss1d}
\end{equation}
and
\begin{multline}
    \kappa_2^{(1)}(\alpha)=-\Phi_\alpha''(0)=8\alpha^{2}
\tau(\alpha)\Big[
      -2(\alpha-1)\alpha(\tau(\alpha)^{2}-1)\tau(\alpha)
      -\frac{2\e^{-2\tau(\alpha)^{2}}\tau(\alpha)}{\pi}
      \\+\frac{
         \e^{-\tau(\alpha)^{2}}\big((4\alpha-2)\tau(\alpha)^{2}-2\alpha+3\big)
       }{\sqrt{\pi}}-\frac{1}{2}\alpha(\alpha-3)
\Big]\ ,
\label{eq:varScaledGauss1d}
\end{multline} 
where $\tau(\alpha)=\operatorname{erf}^{-1}(1-\alpha)$.

   The rate functions from \eqref{eq:legendre} along with numerical simulations are plotted in Fig.~\ref{fig:rateFunction}. 

\end{enumerate}
\end{ex}
\textcolor{black}{Beyond its mathematical interest, the rate function has also concrete practical relevance. As an example, in applications such as portfolio diversification or experimental design, rare fluctuations of the dispersion translate directly into tail risks, and knowing their precise exponential decay rate is operationally valuable.}

\section{Any number of traits ($d\geq 1$) -- Replica approach}\label{sec:replica}

As an alternative route to determine the SCGF of the maximal $M$-dispersion and its cumulants in the limit of large $N$, we adopt the replica approach from the theory of disordered systems~\cite{mezard}.

Let $\bm X_1,\bm X_2,\ldots,\bm X_N$ be independent points in $\R^d$ with common density $f(\bm x)$. 
Define the auxiliary partition function
\begin{equation}\label{eq:def_replica_part_funct_ulam}
   \mathcal{Z}^{(\beta)}_{N,M}(\bm X_1,\ldots,\bm X_N)
   := \sum_{\sigma\in\Sigma_{N,M}}
      \exp\!\left[\frac{\beta}{N}D^M(\bm X_1,\ldots,\bm X_N|\sigma)\right]\ ,
\end{equation}
where the factor $1/N$ in the exponent ensures extensivity of the effective ``internal energy'', and $D^M(\cdot|\sigma)$ is defined in Eq. \eqref{DefMDispersion}. 

Eq.~\eqref{eq:def_replica_part_funct_ulam} is the canonical partition function of a disordered model of $N$ ``constituents'', comprising $M$ spins ($\sigma_i=+1$) and $N-M$ vacancies ($\sigma_i=0$), where the interaction matrix between spins $J_{ij}=|\bm X_i-\bm X_j|^2$ is of the Euclidean type \cite{mezardstructured}. This spin system bears some similarities with the Euclidean matching problems~\cite{sicuro1,sicuro2}, the Sherrington-Kirkpatrick spin glass \cite{mezard}, and the random link travelling salesman problem \cite{TSP}. However -- to our knowledge -- the partition function~\eqref{eq:def_replica_part_funct_ulam} has not been studied previously in this precise form.

At zero temperature $(\beta\to\infty)$, the sum is dominated by configurations that maximise the dispersion,
\begin{equation}\label{eq:replica_lapl_Z_ulam}
   \mathcal{Z}^{(\beta)}_{N,M}
   \approx \exp\!\left[\frac{\beta}{N}D^M_{\max}(\bm X_1,\ldots,\bm X_N)\right]\ ,
\end{equation}
so that
\begin{equation}\label{eq:Dmaxasbetainfty_ulam}
   D^M_{\max}(\bm X_1,\ldots,\bm X_N)
   = \lim_{\beta\to\infty}\frac{N}{\beta}\log\mathcal{Z}^{(\beta)}_{N,M}\ .
\end{equation}
Averaging over the disorder yields the first cumulant
\begin{equation}
   \kappa_1^{(N,M,d)}
   = \lim_{\beta\to\infty}\frac{N}{\beta}\,
     \mathbb{E}\big[\log\mathcal{Z}^{(\beta)}_{N,M}\big]\ .
\end{equation}
The replica identity
\begin{equation}
   \mathbb{E}[\log\mathcal{Z}^{(\beta)}_{N,M}]
   = \lim_{n\to0}\frac{1}{n}    \log\mathbb{E}\big[\left(\mathcal{Z}^{(\beta)}_{N,M}\right)^n\big]\ ,\label{eq:replicaidentityref}
\end{equation}
where $n$ is first promoted to an integer, allows us to evaluate the average over the disorder first. The resulting expression for the scaled first cumulant in the large $N,M$ limit with $M/N=\alpha$ fixed reads
\begin{equation}\label{eq:DmaxAvgReplica_ulam}
   \kappa_1^{(d)}(\alpha)
   = \lim_{\beta\to\infty}\frac{1}{\beta}
     \lim_{n\to0}\frac{1}{n}
     \lim_{N\to\infty}\frac{1}{N}    \log\mathbb{E}\!\left[\big(\mathcal{Z}^{(\beta)}_{N,M}\big)^n\right]\ .
\end{equation}
\textcolor{black}{Note that the replica identity \eqref{eq:replicaidentityref} holds for $N$ finite; in practice, however, the replica method proceeds by first taking $N\to\infty$ at integer $n$, and only afterwards performing the analytic continuation $n\to 0$. This exchange of limits is the standard - though often uncontrolled - step that makes the replica calculation tractable.}
Although the formula \eqref{eq:DmaxAvgReplica_ulam} for the first cumulant can be computed explicitly in all dimensions, it is convenient to follow a different route and determine directly the SCGF $\Phi_\alpha(p)$, which encodes {all} cumulants by differentiation (see Eq. \eqref{eq:leadingcumulantsfromPhi}).

We follow here the recipe proposed in \cite{fyodorov}.  
Evaluating the average of the replicated partition function this time in the double-scaling limit $n\to 0,\beta\to\infty$ with $n\beta=-p=O(1)$ for large $\beta$, one obtains
\begin{align}\label{eq:replica_expec_Z_pb_ulam}
\mathbb{E}\!\left[\big(\mathcal{Z}^{(\beta)}_{N,M}\big)^n\right]
= \mathbb{E}\!\left[
  \exp\!\left[-p\,\frac{1}{\beta}
  \log\mathcal{Z}^{(\beta)}_{N,M}\right]\right] \approx \mathbb{E}\!\left[
  \exp\!\left[-p\,\frac{D^M_{\max}(\bm X_1,\ldots,\bm X_N)}{N}\right]\right],
\end{align}
so that the scaled cumulant generating function can be written as
\begin{align}\label{eq:replica_CGF_ulam}
\Phi_\alpha(p)
= \lim_{N\to\infty}
  -\frac{1}{N}\log
  \mathbb{E}\!\left[\e^{-p\,D^M_{\max}/N}\right]
= \lim_{\beta\to\infty}
   \lim_{N\to\infty}
   -\frac{1}{N}
   \log\mathbb{E}\!\left[
     \big(\mathcal{Z}^{(\beta)}_{N,M}\big)^{-p/\beta}
   \right].
\end{align}
The task is then to evaluate the large-$N$ asymptotics of the disorder-averaged replicated partition function, which we tackle in the next subsection.

\subsection{Averaging the replicated partition function over the disorder}
The expected value of replicated partition function reads explicitly for $n$ integer
\begin{align} \label{eq:replica_avg_Zpb}
\nonumber \mathbb{E} &\left[\left( \mathcal{Z}^{(\beta)}_{N,M}(\bm X_1,\ldots,\bm X_N)\right)^n\right] =  \int_{\mathbb{R}^d} \dots \int_{\mathbb{R}^d} \prod_{i=1}^N \de \bm x_i~f(\bm x_i) 
\sum_{\{\sigma_i^a\}} \prod_{a=1}^n \delta\left(\sum_{i=1}^N \sigma_i^a,M\right) \notag \\
\nonumber &\quad \times\exp\left(\frac{\beta}{N} \sum_{a=1}^n \sum_{i,j}^N |\bm x_i - \bm x_j|^2 \sigma_i^a \sigma_j^a\right)\ \\
&= \sum_{\{\sigma_i^a\}} \dfrac{1}{(2 \pi)^n} \int_0^{2 \pi} \prod_{a=1}^n \de \tau^a \e^{\mathrm{i} \sum_{a=1}^n \tau^a( N \alpha - \sum_{i=1}^N \sigma_i^a)}~\mathbb{E} \left[ \exp\left(\frac{\beta}{N} \sum_{a=1}^n \sum_{i,j=1}^N|\bm X_i-\bm X_j|^2 \sigma_i^a \sigma_j^a\right)\ \right]\ ,
\end{align}
where we used the Fourier representation for the Kronecker $\delta$-functions $\delta(x,y)=\int_0^{2\pi}(\de\tau/(2\pi))\exp(\mathrm{i}\tau(x-y))$, and we used the notation
\begin{equation}
   \sum_{\{\sigma_i^a\}}:= \prod_{k=1}^n \prod_{i=1}^N \sum_{ \sigma^k_i=0,1}=
\sum_{\sigma_1^1=0,1}\cdots \sum_{\sigma_N^1=0,1}\sum_{\sigma_1^2=0,1}\cdots \sum_{\sigma_N^2=0,1}\cdots\sum_{\sigma_N^n=0,1}\ .
\end{equation}
Write the $d$-dimensional vectors in components as $\bm x_i=( 
 x_{i,1},\ldots,  x_{i,d})$, $i=1,\ldots,N$. We now define $n$ mean-field densities, depending on vectors $\bm x=(x_1, \dots, x_d) \in \mathbb{R}^d$, for all $a = 1, \dots,n$, as
\begin{equation} \label{eq:def_mf_densities}
 \rho^a(\bm x) :=\frac{1}{N}\sum_{i=1}^N \delta(\bm  x-\bm 
 x_{i})\sigma_i^a=\frac{1}{N}\sum_{i=1}^N \prod_{\ell=1}^d\delta(x_\ell-x_{i,\ell})\sigma_i^a\ , 
\end{equation}
with normalisation
\begin{equation}
    \int_{\mathbb{R}^d} \de \bm x~\rho^a(\bm x)=\alpha\ .\label{eq:def_mf_densities_norm} 
\end{equation}
From the definition,
\begin{equation}
    \sum_{i,j=1}^N|\bm x_i-\bm x_j|^2\sigma_i^a\sigma_j^a=N^2\int_{\mathbb{R}^d}\int_{\mathbb{R}^d} \de \bm x~\de \bm y~\rho^a(\bm x)\rho^a(\bm y)|\bm x-\bm y|^2\ .
\end{equation}
By enforcing the definition \eqref{eq:def_mf_densities} with functional deltas as
\begin{align}
 \nonumber   1 &=\int\mathcal{D}[\rho^a]\delta\left[N\rho^a(\bm x)-\sum_{i=1}^N \prod_{\ell=1}^d\delta(x_\ell-x_{i,\ell})\sigma_i^a\right]\\
\nonumber    &=\iint\mathcal{D}[\rho^a]\mathcal{D}[\hat\rho^a]\exp\left[\mathrm{i}N\int_{\mathbb{R}^d} \de \bm x~\rho^a(\bm x)\hat\rho^a(\bm x)-\mathrm{i}\int_{\mathbb{R}^d} \de \bm x~\hat\rho^a(\bm x)\sum_{i=1}^N \prod_{\ell=1}^d\delta(x_\ell-x_{i,\ell})\sigma_i^a\right]\\
    &=\iint\mathcal{D}[\rho^a]\mathcal{D}[\hat\rho^a]\exp\left[\mathrm{i}N\int_{\mathbb{R}^d} \de \bm x~\rho^a(\bm x)\hat\rho^a(\bm x)-\mathrm{i}\sum_{i=1}^N \hat\rho^a(\bm x_i)\sigma_i^a\right]\ , 
\end{align}
the expected value \eqref{eq:replica_avg_Zpb} becomes
\begin{align}
 \nonumber   \mathbb{E} &\left[\left( \mathcal{Z}^{(\beta)}_{N,M}(\bm X_1,\ldots,\bm X_N)\right)^n\right]\propto \int\{\mathcal{D}[\rho^a]\}\{\mathcal{D}[\hat\rho^a]\}\de\vec\tau~\e^{\mathrm{i}\alpha N\sum_a\tau_a}\\
\nonumber    &\times \exp\left[\mathrm{i}N\sum_a\int_{\mathbb{R}^d} \de\bm x~\rho^a(\bm x)\hat\rho^a(\bm x)+\beta N\sum_a \int_{\mathbb{R}^d}\int_{\mathbb{R}^d} \de\bm x~\de\bm y~\rho^a(\bm x)\rho^a(\bm y)|\bm x - \bm y|^2\right]\\ &\times \sum_{\{\sigma_i^a\}}
    \mathbb{E}\left[ \exp\left(-\mathrm{i} \sum_{a=1}^n \sum_{i=1}^N \hat{\rho}^a(\bm X_i) \sigma_i^a -\mathrm{i} \sum_{a=1}^n \tau^a \sum_{i=1}^N \sigma_i^a \right)\right]\ ,
\end{align}
where we omitted proportionality constants, and denoted   $\{\mathcal{D}[\rho^a]\} := \prod_{a=1}^n  \mathcal{D}[\rho^a]$, $\{\mathcal{D}[\hat\rho^a]\} := \prod_{a=1}^n  \mathcal{D}[\hat{\rho}^a]$, and $\de\vec\tau := \prod_{a=1}^n \de \tau^a$.

Now 
\begin{align}
  \nonumber &\sum_{\{\sigma_i^a\}} \mathbb{E}\left[\exp\left(-\mathrm{i} \sum_{a=1}^n \sum_{i=1}^N \hat{\rho}^a(\bm X_i) \sigma_i^a -\mathrm{i} \sum_{a=1}^n \tau^a \sum_{i=1}^N \sigma_i^a \right)\right]= \\ 
  &= \prod_{i=1}^N \left[ \int_{\mathbb{R}^d} \de\bm x_i f(\bm x_i)  \prod_{a=1}^n\sum_{\sigma_i^a=0,1} (\e^{-\mathrm{i} \sigma_i^a \hat{\rho}^a(\bm x_i)-\mathrm{i} \sigma_i^a \tau^a})  \right] = \left[  \int_{\mathbb{R}^d} \de\bm x f(\bm x) \prod_{a=1}^n (1+ \e^{-\mathrm{i} \hat{\rho}^a(\bm x)-\mathrm{i}\tau^a})  \right]^N\ .
\end{align}
Therefore, the average of the replicated partition function can be cast in the form
\begin{align}
   \mathbb{E} &\left[\left( \mathcal{Z}^{(\beta)}_{N,M}(\bm X_1,\ldots,\bm X_N)\right)^n\right] &\propto \int\{\mathcal{D}[\rho^a]\}\{\mathcal{D}[\hat\rho^a]\}\de\vec\tau~\e^{N\mathcal{S}^{(\beta,n)}[\{\rho^a\},\{\hat{\rho}^a\},\{\tau^a\},\{\lambda^a\}]}\ ,
\end{align}
where the action $\mathcal{S}^{(\beta,n)}$ is defined as
\begin{align}
\nonumber   &\mathcal{S}^{(\beta,n)}[\{\rho^a\},\{\hat{\rho}^a\},\{\tau^a\},\{\lambda^a\}]:= \beta \sum_{a=1}^n \int_{\mathbb{R}^d}\int_{\mathbb{R}^d} \de\bm x~\de\bm y~\rho^a(\bm x)\rho^a(\bm y)|\bm x - \bm y|^2 \\
\nonumber &+ \mathrm{i}\sum_{a=1}^n\int_{\mathbb{R}^d} \de\bm x~\rho^a(\bm x)\hat\rho^a(\bm x)+  \mathrm{i} \alpha \sum_{a=1}^n \tau^a + \log{\int_{\mathbb{R}^d}\de \bm x~ f(\bm x) \e^{\sum_{a=1}^n \log{(1+\e^{-\mathrm{i}\hat{\rho}^a(\bm x)-\mathrm{i} \tau^a}})}} \\
&- \sum_{a=1}^n \lambda^a\left(\int_{\mathbb{R}^d} \de \bm x \rho^a(\bm x) - \alpha\right)\ .\label{actionmathcalS}
\end{align}
The Lagrange multipliers $\{\lambda^a\}$ enforce the normalisation of the mean-field densities \eqref{eq:def_mf_densities_norm}. The form \eqref{actionmathcalS} lends itself to a saddle-point evaluation for large $N$, leading to
\begin{equation}\label{eq:replica_N_inf_replicatedZavg}
   \mathbb{E} \left[\left( \mathcal{Z}^{(\beta)}_{N,M}(\bm X_1,\ldots,\bm X_N)\right)^n\right] 
   \overset{N \to \infty}{\approx} \e^{N\mathcal{S}^{(\beta,n)}[\{\rho*^a\}, \{\hat{\rho}*^a\},\{t*^a\},\{\lambda*^a\}]}\ ,
\end{equation}
where $[\{\rho*^a\}, \{\hat{\rho}*^a\},\{t*^a\},\{\lambda*^a\}]$ maximise $\mathcal{S}^{(\beta,n)}$.
The saddle-point equations read $\forall \ a = 1,...,n$ (omitting the $*$ for simplicity),
\begin{align}
    \frac{\delta\mathcal{S}^{(\beta,n)}}{\delta \rho^a}=0 &\Rightarrow \mathrm{i}\hat\rho^a(\bm x)+2\beta\int_{\mathbb{R}^d} \de \bm y~\rho^a(\bm y)|\bm x-\bm y|^2-\lambda^a=0\ , \label{SP1}\\
    \frac{\delta\mathcal{S}^{(\beta,n)}}{\delta \hat\rho^a}=0 &\Rightarrow\mathrm{i}\rho^a(\bm x)+\frac{f(\bm x) \e^{\sum_b\log(1+\e^{-\mathrm{i}\hat\rho^b(\bm x)-\mathrm{i}\tau^b})}\left[\frac{\e^{-\mathrm{i}\hat\rho^a(\bm x)-\mathrm{i}\tau^a}(-\mathrm{i})}{1+\e^{-\mathrm{i}\hat\rho^a(\bm x)-\mathrm{i}\tau^a}}\right]}{\int_{\mathbb{R}^d} \de\bm y~f(\bm y)\e^{\sum_b\log(1+\e^{-\mathrm{i}\hat\rho^b(\bm y)-\mathrm{i}\tau^b})}}=0\ ,\label{SP2}\\
    \frac{\partial\mathcal{S}^{(\beta,n)}}{\partial \tau^a}=0 &\Rightarrow \mathrm{i}\alpha+\frac{\int_{\mathbb{R}^d} \de\bm y~f(\bm y) \e^{\sum_b\log(1+\e^{-\mathrm{i}\hat\rho^b(\bm y)-\mathrm{i}\tau^b})}\left[\frac{\e^{-\mathrm{i}\hat\rho^a(\bm y)-\mathrm{i}\tau^a}(-\mathrm{i})}{1+\e^{-\mathrm{i}\hat\rho^a(\bm y)-\mathrm{i}\tau^a}}\right]}{\int_{\mathbb{R}^d} \de\bm y~f(\bm y)\e^{\sum_b\log(1+\e^{-\mathrm{i}\hat\rho^b(\bm y)-\mathrm{i}\tau^b})}}=0\ .
\end{align}
From \eqref{SP1} we get
\begin{equation}
    -\mathrm{i} \hat{\rho}^a(x_1, \dots,x_d) = 2 \beta \left[ \alpha\sum_{\ell=1}^d x_\ell^2  + \overline{(r^2)^a}-2 \sum_{\ell=1}^d x_\ell \overline{(x_\ell)^a} \right] - \lambda^a\ ,\label{minusihatrhoLD}
\end{equation}
where we defined
\begin{align}
   \overline{(r^2)^a} &:= \int_{\mathbb{R}^d} \de x_1 \cdots \de x_d \ \rho^a(x_1, \dots,x_d) \sum_{\ell=1}^d x_\ell^2  \ ,\label{r2bara}\\
   \overline{(x_\ell)^a} &:= \int_{\mathbb{R}^d}\de x_1 \cdots \de x_d ~ \rho^a(x_1, \dots,x_d)~x_\ell \ ,\qquad \forall~\ell=1,\dots,d ,\label{xbara}
\end{align}
and we used the normalisation condition \eqref{eq:def_mf_densities_norm}.

From \eqref{SP2} we obtain 
\begin{align}
    \rho^a(\bm x) = \frac{f(\bm x) \e^{\sum_b\log(1+\e^{-\mathrm{i}\hat\rho^b(\bm x)-\mathrm{i}\tau^b})}\left[\frac{\e^{-\mathrm{i}\hat\rho^a(\bm x)-\mathrm{i}\tau^a}}{1+\e^{-\mathrm{i}\hat\rho^a(\bm x)-\mathrm{i}\tau^a}}\right]}{\int \de\bm y f(\bm y)\e^{\sum_b\log(1+\e^{-\mathrm{i}\hat\rho^b(\bm y)-\mathrm{i}\tau^b})}}\ . \label{eq:rho}
\end{align}
We now assume a \emph{replica-symmetric}  structure, namely there should not be any dependence on the specific replica index $a$ chosen (see \cite{mezard,castellani,zamponi,marinari} for detailed discussions about replica symmetry and its breakdown in disordered spin models)
\begin{equation}
    \rho^a \equiv \rho \, ; \hat{\rho}^a \equiv \hat{\rho} ; \tau^a \equiv \tau \, ; \lambda^a \equiv \lambda\ .
\end{equation}
Equation \eqref{eq:rho} therefore becomes
\begin{equation}
    \rho(\bm x) = \frac{f(\bm x) \e^{n\log(1+\e^{-\mathrm{i}\hat\rho(\bm x)-\mathrm{i}\tau})}\left[\frac{\e^{-\mathrm{i}\hat\rho(\bm x)-\mathrm{i}\tau}}{1+\e^{-\mathrm{i}\hat\rho(\bm x)-\mathrm{i}\tau}}\right]}{\int \de\bm y f(\bm y)\e^{n\log(1+\e^{-\mathrm{i}\hat\rho(\bm y)-\mathrm{i}\tau})}}\ .
\end{equation}
Inserting now Eq. \eqref{minusihatrhoLD} for $\mathrm{i}\hat\rho$, we get
\begin{equation}
  \rho(\bm x) = \frac{f(\bm x) \e^{n\log(1+\e^{\beta\xi(\bm x)})}\left[\frac{\e^{\beta\xi(\bm x)}}{1+\e^{\beta\xi(\bm x)}}\right]}{\int \de\bm y f(\bm y)\e^{n\log(1+\e^{\beta\xi(\bm y)})}}\ ,\label{rhonbetaLD}
\end{equation}
where
\begin{equation}
    \xi(\bm x)=2\left[\alpha\sum_{\ell=1}^d x_\ell^2+\overline{r^2}-2\sum_{\ell=1}^d x_\ell \overline{x_\ell} \right]-\eta=\xi_0(\bm x)-\eta
\end{equation}
with $\eta=\lambda+\mathrm{i}\tau$ (after a harmless rescaling $\lambda\to\beta\lambda$ and similarly for $\mathrm{i}\tau$). 

Multiplying \eqref{SP1} by $\rho(\bm x)$ and integrating we obtain
\begin{align}
    &\beta\iint \de\bm x \de\bm y~\rho(\bm x)\rho(\bm y)|\bm x-\bm y|^2 =\frac{\lambda\alpha\beta}{2}-\frac{\mathrm{i}}{2}\int \de\bm x \rho(\bm x)\hat\rho(\bm x) =\\
    &=\frac{\lambda\alpha\beta}{2}+\frac{\beta}{2}\int \de\bm x~\rho(\bm x)[\xi_0(\bm x)-\lambda]=\frac{\beta}{2}\int \de\bm x~\rho(\bm x)\xi_0(\bm x)\ ,
\end{align}
where we used again the normalisation condition \eqref{eq:def_mf_densities_norm} and \eqref{minusihatrhoLD}.

Inserting this back into \eqref{actionmathcalS} within the replica-symmetry assumption, we get for the action at the saddle point

\begin{align}
\nonumber\mathcal{S}^{(\beta,n)}[\rho;\hat{\rho};\eta] &= \frac{n\beta}{2}\int \de\bm x~\rho(\bm x)\xi_0(\bm x)+n\beta\int \de\bm x~\rho(\bm x)[\lambda-\xi_0(\bm x)]+\mathrm{i\alpha}\beta n\tau\\
\nonumber &+\log \int \de\bm x~f(\bm x)\e^{n\log (1+\e^{\beta\xi(\bm x)})}\\
&=-\frac{n\beta}{2}\int \de\bm x~\rho(\bm x)\xi_0(\bm x)+\alpha\beta n\eta+\log \int \de\bm x~f(\bm x)\e^{n\log (1+\e^{\beta\xi(\bm x)})}\ .\label{actionSLDrestorednbeta}
\end{align}

\subsection{SCGF of the maximal $M$-dispersion}\hfill\\
Now we can set $n = -p/\beta$ in both the saddle-point density and the action, and extract the $\beta\to\infty$ limit to compute the SCGF using Eq. \eqref{eq:replica_CGF_ulam}.

Starting with the density in \eqref{rhonbetaLD}, and using the limits 
\begin{align} \label{eq:xi_limitlog}
    \lim_{\beta\to\infty}\frac{1}{\beta}\log(1+\e^{\beta\xi}) &=\xi\theta(\xi) \ ,\\ \label{eq:xi_limitfrac}
    \lim_{\beta\to\infty}\frac{\e^{\beta\xi}}{1+\e^{\beta\xi}} &=\theta(\xi)\ ,
\end{align}
where $\theta(x)$ is the Heaviside step function, we have
\begin{equation}
    \rho(\bm x)\sim
    \frac{f(\bm x)\e^{-p\xi(\bm x)\theta(\xi(\bm x))}\theta(\xi(\bm x))}{\int \de\bm y~f(\bm y)\e^{-p\xi(\bm y)\theta(\xi(\bm y))}}\label{densityLDscalinglimit}
\end{equation}
for large $\beta$. The term in the exponent can be simplified as
\begin{equation}
   \e^{-p\xi(\bm x)\theta(\xi(\bm x))}= \theta(\xi(\bm x))\e^{-p\xi(\bm x)}+1-\theta(\xi(\bm x))\ .
\end{equation}
Inserting the density \eqref{densityLDscalinglimit} into the action \eqref{actionSLDrestorednbeta}, we have to compute the following terms:
\begin{align}
\nonumber    \int \de\bm x~\rho(\bm x)\xi_0(\bm x) &=\frac{\int \de\bm x f(\bm x)\theta(\xi(\bm x))\xi_0(\bm x)[\theta(\xi(\bm x))\e^{-p\xi(\bm x)}+1-\theta(\xi(\bm x))]}{\int \de\bm x~f(\bm x)[\theta(\xi(\bm x))\e^{-p\xi(\bm x)}+1-\theta(\xi(\bm x))]}\\
&=\frac{\int_\Gamma \de\bm x f(\bm x)\e^{-p\xi(\bm x)}\xi_0(\bm x)}{\int_\Gamma \de\bm x f(\bm x)[\e^{-p\xi(\bm x)}-1]+1}\label{firsttermrescaledSLD}\\
\nonumber\int \de\bm x~f(\bm x)\e^{n\log (1+\e^{\beta\xi(\bm x)})} &\equiv \int \de\bm x~f(\bm x)\e^{-\frac{p}{\beta}\log (1+\e^{\beta\xi(\bm x)})} \sim \int_\Gamma \de\bm x f(\bm x)\e^{-p\xi(\bm x)}+1\\
&-\int_\Gamma \de\bm x~f(\bm x)\ ,\label{secondtermrescaledSLD}
\end{align}
where the domain of integration $\Gamma$ is such that $\xi(\bm x)>0$, i.e.
\begin{equation}
    \Gamma=\left\{(x_1, \dots,x_d)\in\mathbb{R}^d: 2\left[\alpha \sum_{i=1}^dx_i^2+\overline{r^2}-2\sum_{i=1}^dx_i\overline{x}\right]-\eta>0\right\}\ .
\end{equation}
The constants $\overline{r^2},\overline{x_i},\eta$ are determined by the $d+2$ conditions 
\begin{align}
    \alpha &=\frac{\int_\Gamma \de\bm x f(\bm x)\e^{-p\xi(\bm x)}}{\int_\Gamma \de\bm x f(\bm x)[\e^{-p\xi(\bm x)}-1]+1}\label{alphanormLD}\\
    \overline{x_i} &=\frac{\int_\Gamma \de\bm x f(\bm x)\e^{-p\xi(\bm x)}x_i}{\int_\Gamma \de\bm x f(\bm x)[\e^{-p\xi(\bm x)}-1]+1}\label{overlinexLD} \ , \text{for all $ i =1,\dots,d$}\\
          \overline{r^2} &=\frac{\int_\Gamma \de\bm x f(\bm x)\e^{-p\xi(\bm x)}\sum_{i=1}^d x_i^2}{\int_\Gamma \de\bm x f(\bm x)[\e^{-p\xi(\bm x)}-1]+1}\ ,\label{rsquareLD}
\end{align}
which follows from Eqs. \eqref{eq:def_mf_densities_norm}, \eqref{r2bara} and \eqref{xbara} with the replica index $a$ suppressed, and $\rho(\bm x)$ given by \eqref{densityLDscalinglimit}.

Combining Eqs. \eqref{eq:replica_CGF_ulam}, \eqref{eq:replica_N_inf_replicatedZavg}, \eqref{actionSLDrestorednbeta}, \eqref{firsttermrescaledSLD} and \eqref{secondtermrescaledSLD}, and assuming replica symmetry, we conclude that
\begin{equation}
   \mathbb{E}\!\left[
  \exp\!\left[-p\,\frac{D^M_{\max}(\bm X_1,\ldots,\bm X_N)}{N}\right]\right]\approx \mathrm{e}^{- N \Phi_\alpha (p)}\ ,\label{MGFconnectedwithratePhialpha}
\end{equation}
where the cumulant generating function $\Phi_\alpha (p)$ is 
\begin{equation}
\Phi_\alpha (p)=-\frac{p}{2}\frac{\int_\Gamma \de\bm x f(\bm x)\e^{-p\xi(\bm x)}\xi_0(\bm x)}{\int_\Gamma \de\bm x f(\bm x)[\e^{-p\xi(\bm x)}-1]+1}+\alpha p \eta -\log\left[\int_\Gamma \de\bm x f(\bm x)\e^{-p\xi(\bm x)}+1-\int_\Gamma \de\bm x~f(\bm x)\right]\ .  \label{ratePhiLaplaceLD}   
\end{equation}
Note that in this replica derivation we did not assume rotational invariance of the probability density $f(\bm x)$. If we restore this condition, we are now able to show that \eqref{ratePhiLaplaceLD} -- supplemented by the conditions \eqref{alphanormLD} and \eqref{rsquareLD} -- precisely recovers the formula \eqref{eq:SCGF_radial} obtained from order statistics considerations. This proof will therefore conclude our alternative derivation of the main formula for the SCGF of the maximal $M$-dispersion problem in $d$ dimensions. 

For a rotationally symmetric $f(\bm x)$ (see~\eqref{eq:fbmxrotinvdensity}), the integration domain $\Gamma$ simplifies to
\begin{equation}
    \Gamma=\{(x_1, \dots,x_d)\in\mathbb{R}^d: \xi_0(\bm x)-\eta>0\}\label{GammadefAppendix}
\end{equation}
where
\begin{equation}
    \xi_0(\bm x)=2\left[\alpha \sum_{\ell=1}^d x_\ell^2+\overline{r^2}\right]\ .
\end{equation}
The condition in \eqref{GammadefAppendix} implies that the radial integrals should be restricted to the domain $r>R=\sqrt{(\eta-2\overline{r^2})/(2\alpha)}$.

Therefore, using \eqref{alphanormLD} and \eqref{rsquareLD}
\begin{align}
   \frac{\int_\Gamma \de\bm x f(\bm x)\e^{-p\xi(\bm x)}\xi_0(\bm x)}{\int_\Gamma \de\bm x f(\bm x)[\e^{-p\xi(\bm x)}-1]+1} &=\frac{\int_\Gamma \de\bm x f(\bm x)\e^{-p\xi(\bm x)}\left(2\alpha |\bm x|^2+2\overline{r^2}\right)}{\int_\Gamma \de\bm x f(\bm x)[\e^{-p\xi(\bm x)}-1]+1}=4\alpha \overline{r^2}\\
   \int_\Gamma \de\bm x f(\bm x)\e^{-p\xi(\bm x)}+1-\int_\Gamma \de\bm x~f(\bm x) &=\frac{1}{\alpha}\int\de\bm x f(\bm x)\e^{-p\xi(\bm x)}\ .
\end{align}
After simplifications, Eq. \eqref{ratePhiLaplaceLD} turns into 
\begin{align}
\nonumber\Phi_\alpha (p) &=2\alpha^2 p R^2-\log\left[\frac{1}{\alpha}\int\de\bm x f(\bm x)\e^{-p\xi(\bm x)}\right]\\
\nonumber &=2\alpha^2 p R^2-\log\left[\frac{\e^{2\alpha p R^2}}{\alpha}\int_R^\infty\de r g(r)r^{d-1}\e^{-2\alpha p r^2}\right]\\
&=-(1-\alpha)(2\alpha p R^2)-\log\left[\frac{1}{\alpha}\int_R^\infty\de r g(r)r^{d-1}\e^{-2\alpha p r^2}\right]\ ,\label{PhialphapAppendix}
\end{align}
where $R$ satisfies the equation \eqref{alphanormLD} in the form
\begin{equation}
 \alpha =\frac{\int_\Gamma \de\bm x f(\bm x)\e^{-p\xi(\bm x)}}{\int_\Gamma \de\bm x f(\bm x)[\e^{-p\xi(\bm x)}-1]+1}=\frac{\int_R^\infty \de r~r^{d-1}g(r)\e^{-p[2\alpha r^2+2\overline{r^2}-\eta]}}{\int_R^\infty \de r~r^{d-1}g(r)\e^{-p[2\alpha r^2+2\overline{r^2}-\eta]} +\int_0^R\de r~g(r)r^{d-1}}\ ,\label{alphanormLDappendix}    
\end{equation}
 where we used the normalisation $\int_\Gamma\de\bm x f(\bm x)+\int_{\overline{\Gamma}}\de\bm x f(\bm x)=1$ by normalisation, with $\overline{\Gamma}$  the complement of $\Gamma$.

Inverting the equation, and using $2\overline{r^2}-\eta=-2\alpha R^2$ we get
\begin{equation}
    \frac{1}{\alpha}=1+\frac{\int_0^R\de r~g(r)r^{d-1}}{\e^{2\alpha p R^2}\int_R^\infty \de r~r^{d-1}g(r)\e^{-2\alpha p r^2}}\ ,\label{EqforRappendix}
\end{equation}
which coincides with Eq. \eqref{eq:R_sol}, obtained from order statistics considerations. 

Extracting the term $2\alpha p R^2$ from \eqref{EqforRappendix}, we get
\begin{equation}
   2\alpha p R^2=-\log\frac{1-\alpha}{\alpha}+\log\int_0^R\de r~g(r)r^{d-1}-\log\int_R^\infty\de r~r^{d-1}g(r)\e^{-2\alpha p r^2}\ .
\end{equation}
Inserting this expression into \eqref{PhialphapAppendix}, we rewrite the SCGF as
\begin{align}
    \Phi_\alpha(p)=- \alpha\log\left(\frac{1}{\alpha}\int_{r\geq R}r^{d-1}g(r)\e^{-2\alpha p r^2} \de r\right)
   -(1- \alpha)\log\left(\frac{1}{1-\alpha}\int_{r\leq R}r^{d-1}g(r)\de r\right)\ ,
\end{align}
which indeed coincides with Eqs.~\eqref{eq:aux_SCGF_radial} and~\eqref{eq:SCGF_radial}. This concludes the proof of equivalence between the two approaches.

\section{Numerical algorithm} 
\label{sec:heuristics}
The numerical method used to perform the simulations in dimension $d>1$ is the greedy \emph{constructive heuristic C-2} presented in \cite{gloverHeuristics}. Arranging the $N$ points $\bm{x}_i\in\R^d$ as the columns of a matrix $A\in\R^{d\times N}$, the task is to select $M$ columns from $A$ such that \eqref{DefMDispersion} is maximized. Denoting with $S$ the set of the selected indexes $(|S|=M)$, C-2 builds $S$ from the empty set by repeatedly adding the element $\bm{x}_i$ that maximizes the
\emph{composite squared distances} to a reference set $X$, defined as
\begin{equation}
	D(\bm{x}_i,X)=\sum_{\bm{x}_j\in X} |\bm{x}_i- \bm{x}_j|^2\ .
\end{equation} 
The squared distances $D_{ij}=|\bm{x}_i-\bm{x}_j|^2$ for all $i,j$  can be computed efficiently in a time $O(dN^2)$ via the identity
\begin{equation}
    D_{ij}=|\bm{x}_i|^2+|\bm{x}_j|^2-2 (A^T A)_{ij}\ .
\end{equation}
The greedy routine then essentially goes as follows:
\begin{itemize}
  \item \textbf{Initialisation:} choose the first column whose total distance to all columns is largest, i.e.,
  $\arg\max_i \sum_{j=1}^N D_{ij}$, corresponding to C-2's first selection using $X=S$ (the full set) as the reference.
  \item \textbf{Iteration:} maintain a running \emph{score} for each candidate $j\notin S$,
  \[
  \mathrm{score}(j) \;=\; \sum_{i\in S} D_{ij}\ ,
  \]
  and add the $j$ with maximum score. This is exactly the C-2 composite distance criterion with $X$ taken as the current selected set.
\end{itemize}
A pseudocode of the algorithm described above is included in Table~\ref{table:Alg}. 

\begin{figure}
    \centering
    \includegraphics[width=0.99\linewidth]{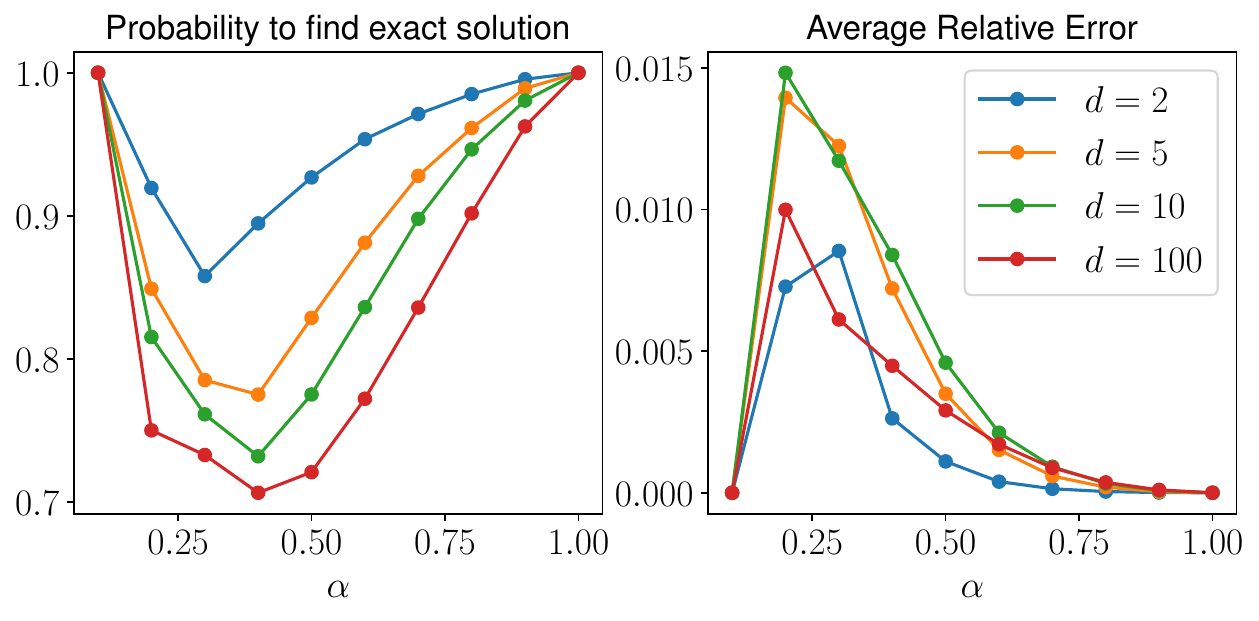}
    \caption{Comparison between the results obtained from the heuristic algorithm and the exact solution obtained via a brute-force approach in $N=10$, for several values of $\alpha=M/N$ and $d$. The sample-size for the estimation of probability and averages is $10^3$. (Left) Probability that the set selected by the greedy algorithm matches the exact solution from a brute-force algorithm. (Right) Average relative error in the maximal dispersion, computed as $(D^M_{\mathrm{greedy}}-D^M_{\mathrm{exact}})/D^M_{\mathrm{exact}}$.}
    \label{fig:Benchmark}
\end{figure}

For a small value of $N$, the performances of this heuristic have been compared with the exact solutions obtained via a brute force approach, for several values of the dimension of the space $d$ and $\alpha=M/N$. The results are shown in Figure~\ref{fig:Benchmark}. 
\begin{table}[t]
\centering
\begin{tabular}{|p{0.97\linewidth}|}
\hline
\begin{minipage}{0.97\linewidth}
\vspace{5pt}
\textbf{Algorithm 1: Greedy algorithm for the maximum dispersion}\\
\vspace{-15pt}
\begin{algorithmic}[1]
\Require $A\in\mathbb{R}^{d\times N}$ with columns $\bm{x}_1,\dots,\bm{x}_N$; target size $M\le N$
\Ensure Selected indices $S$; dispersion value $\mathrm{Dis}(S)$
\State Compute $n_i\gets | \bm{x}_i|^2$ for $i=1,\dots,N$; compute Gram matrix $G\gets A^T A$
\State Build squared-distance matrix $D$ with $D_{ij}\gets n_i+n_j-2G_{ij}$ for all $i,j$
\State $i_1 \gets \arg\max_{i\in\{1,\dots,N\}} \sum_{j=1}^N D_{ij}$ \Comment{largest total distance}
\State $S\gets \{i_1\}$
\State $\mathrm{score}(j)\gets D_{j,i_1}$ for all $j\neq i_1$; set $\mathrm{score}(i_1)\gets -\infty$
\For{$t=2$ to $M$}
  \State $j^\star \gets \arg\max_{j\notin S}\ \mathrm{score}(j)$
  \State $S\gets S\cup\{j^\star\}$
  \State $\mathrm{score}(j)\gets \mathrm{score}(j)+D_{j,j^\star}$ for all $j$
  \State $\mathrm{score}(j)\gets -\infty$ for all $j\in S$ \Comment{prevent reselection}
\EndFor
\State $\mathrm{Dis}(S)\gets \sum_{i\in S}\sum_{j\in S} D_{ij}$ \Comment{equals $\sum_{i<j}D_{ij}$}
\State \Return $S,\ \mathrm{Dis}(S)$
\vspace{10pt}
\end{algorithmic}
\end{minipage}\\
\hline
\end{tabular}
\caption{A greedy constructive heuristic for the maximum dispersion problem, mirroring the C-2 constructive heuristics described in~\cite{gloverHeuristics}.}
\label{table:Alg}
\end{table}

\section{Conclusion and Outlook}
\label{sec:conclusion}

In this paper, we considered the full statistics of the maximal $M$-dispersion of $N$ random points in $d$ dimensions. This problem originates in the field of operations research, where the objective is to select a ``sub-committee'' of $M$ members out of a population of $N$ individuals each having $d$ different traits, in such a way that the traits of the sub-committee's members are as ``dispersed'' as possible. This is in general a difficult problem that -- for large but fixed instances -- can only be solved approximately using heuristic algorithms whose performance is often tested on \emph{randomly generated} instances \cite{gallo1980}. However, the current lack of analytical control over the expected maximal dispersion and its fluctuations in these random instances makes it difficult to establish a reliable benchmark to assess the performance of such heuristic algorithms. 

In our work, we use an additive measure of dispersion that receives larger contributions whenever the pairwise distance between the trait vectors increases (see Eq. \eqref{DefMDispersion}). First, for $d=1$ we could characterise exactly the ``geometry'' of the optimising subset: on the line, one should always pick the $k$ leftmost and $M-k$ rightmost points, for an optimal $k$ to be determined instance by instance (see Section \ref{sec:order_stat}). Guided by this intuition, we set up a mean-field theory valid for $d\geq 1$ and $M,N$ large whereby the set of points that typically maximise dispersion would lie on the complementary of a ball centered at the origin, and whose radius could be precisely characterised (Sec. \ref{subsec:typical}). With a slight tilting of the argument, we could also extract the rate (or large deviation) function governing rare fluctuations of the maximal dispersion (Sec. \ref{subsec:SCGF}). Alternatively, the same final formulae for the SCGF and the rate function are obtained from a replica approach, where the maximiser of the dispersion is obtained from the zero-temperature free energy corresponding to an auxiliary thermodynamical system in the canonical ensemble (Sec. \ref{sec:replica}).  

For the special case of a single trait ($d=1$), we exploited exact tools from order statistics -- the field of probability theory dealing with random variables that are sorted in increasing order -- to compute average and variance of the dispersion of \emph{balanced} configurations for finite $N,M$ (Sec. \ref{sec:backfiniteN}). Balanced configurations are prefix-suffix with the same number of points in the two extreme blocks: while these configurations are asymptotically optimal for large instances, they do not necessarily yield the maximal dispersion on every single, finite instance. However, they provide excellent approximants even for moderate values of $N$, while allowing an exact finite-$N,M$ treatment.

Using exact enumerations for smaller instances in $d=1$ and heuristic numerical algorithms for larger dimension, we could show excellent agreement between simulations and our theoretical predictions. As shown in Fig.~\ref{fig:rateFunction}, the maximal $M$-diversity has Gaussian  fluctuations around the mean on a scale $\sim O(1/\sqrt{N})$, whereas larger fluctuations $O(1)$ are governed by the full rate functions computed here.

One of the future challenges will be to consider a dispersion measure where the objective function to maximise is actually more sensitive to the \emph{local} gaps between traits: consider for instance the quadratic $M$-dispersion measure defined in Eq. \eqref{DefMDispersion} for the $d=1$ case, with the single trait being the \emph{wealth} of each individual. The subset of a large population whose wealth is as dispersed as possible would be formed by selecting the $\sim M/2$ poorest and $\sim M/2$ richest individuals in the population, which leads to a sub-committee not necessarily representative of all intermediate classes of wealth. \textcolor{black}{For instance, maximising the \emph{minimum} pairwise distance among the selected individuals---rather than their \emph{sum} as in Eq. \eqref{DefMDispersion}---would prevent the optimal subset from clumping at the boundary of the distribution, and may be more natural in applications requiring more uniform coverage of the trait space. The additive measure studied here should therefore be viewed as a first starting point within a broader family of dispersion problems.} Another interesting direction of research will be to consider a case where the $N$ individuals in the initial population are not created equal, or where there are correlations or particular structures between traits across different individuals that may better mimic the complexity and ``hardness'' of real-life scenarios. Moreover, the full analysis of the finite $N,M$ case in all dimensions is still a challenging open problem. Finally, the mean-field approach developed in Sec. \ref{sec:continuous} hinges on the pdf $f(\bm x)$ being sufficiently light-tailed: it would be very interesting to find a corresponding theory for heavy-tailed trait distributions as well.

\appendix

\section{Generalities on order statistics $(d=1)$} \label{app:order_stat_generalities}

Let $X_1,\ldots,X_N$ be i.i.d. real random variables with common distribution function $F(x)$ with density  $f(x)$. We denote by $ X_{(1)}, \ldots,X_{(N)} $ their order statistics, $X_{(1)} \leq X_{(2)} \leq \cdots \leq X_{(N)}$. 

From the general discussion in the main text, the maximal $M$-dispersion will be the sum of the quadratic difference between the $i$ smallest and the $N-j+1$ largest  order statistics on the line, for some $1\leq i<j\leq N$. 
For all $i<j$, the joint density of the first $i$ and the last $N-j+1$ order statistics $ X_{(1)}, \cdots,X_{(i)},X_{(j)},  \cdots,X_{(N)}$ is
\begin{align}
\nonumber & f_{X_{(1)},\ldots,X_{(i)},X_{(j)},\ldots,X_{(N)}}(x_1,\ldots,x_i,x_j,\ldots,x_N)=\\
&\frac{N!}{(j-i-1)!}f(x_1)\cdots f(x_i)\left[F(x_j)-F(x_i)\right]^{j-i-1}f(x_j)\cdots f(x_N) \prod_{i=1}^{N-1} \theta(x_{i+1}-x_i)\ ,\label{eq:jointfirstiandlastj}
\end{align}
where the Heaviside function is $\theta(x)=1$ for all $x \ge 0$ and $\theta(x)=0$ otherwise. The interpretation of \eqref{eq:jointfirstiandlastj} is textbook:  there are random variables taking values at $x_1,\cdots,x_i, x_j,\cdots,x_N$ (which happens according to the respective probability densities), multiplied by the probability that all other $j-i-1$ variables sit between $x_i$ and $x_j$; the combinatorial factor counts the number of ways the variables can be distributed in the three groups (left-middle-right).

Similarly, the bivariate density of $X_{(i)},X_{(j)}$ is
\begin{align} 
\nonumber & f_{X_{(i)},X_{(j)}}(x_i,x_j)=\frac{N!}{(i-1)!(j-i-1)!(N-j)!}\\
 &\times F(x_i)^{i-1}f(x_i)\left[F(x_j)-F(x_i)\right]^{j-i-1}f(x_j) (1-F(x_j))^{N-j}\theta(x_j-x_i)\ . 
\end{align}
It follows that the conditional density of $X_{(1)},\ldots,X_{(i)},X_{(j)},\ldots,X_{(N)}$ conditioned on $X_{(i)},X_{(j)}$ is
\begin{align} 
\nonumber & f_{X_{(1)},\ldots,X_{(i)},X_{(j)},\ldots,X_{(n)}|X_{(i)},X_{(j)}}(x_1,\ldots,x_i,x_j,\ldots,x_N)=\\
\nonumber & \frac{f_{X_{(1)},\ldots,X_{(i)},X_{(j)},\ldots,X_{(n)}}(x_1,\ldots,x_i,x_j,\ldots,x_N)}{f_{X_{(i)},X_{(j)}}(x_i,x_j)}=\\
&=(i-1)!(N-j)!\frac{f(x_1)}{F(x_i)}\cdots \frac{f(x_{i-1})}{F(x_i)}\frac{f(x_{j+1})}{1-F(x_j)}\cdots \frac{f(x_N)}{1-F(x_j)}\prod_{i=1}^{N-1} \theta(x_{i+1}-x_i)\ . \label{eq:1d_orderstat_conditioned}
\end{align}
From \eqref{eq:1d_orderstat_conditioned} we see that $X_{(1)},\ldots,X_{(i)},X_{(j)},\ldots,X_{(n)}$ conditioned on $X_{(i)},X_{(j)}$ have the same density of the union of two order statistics:
\begin{enumerate}
    \item the order statistics of a set of $(i-1)$ i.i.d. random variables in the (random) interval $(-\infty, X_{(i)}]$ with truncated and renormalised density $\frac{f(x)}{F(X_{(i)})}\theta(X_{(i)}-x) $, and 
    \item the order statistics of a set of $(N-j)$ i.i.d. random variables in the (random) interval $[X_{(j)},+\infty)$ with truncated and renormalised density $\frac{f(x)}{1-F(X_{(j)})}\theta(x-X_{(j)})$.
\end{enumerate}
The conditional density in \eqref{eq:1d_orderstat_conditioned} precisely features in the general expectation formula \eqref{eqmaintext:expgbalanced} of the main text. For more details, we refer the reader to the classical reference~\cite{OrderStatistics}.

\section{Function $H_M(a,b)$} \label{appendix:H}
We provide the expression of the function $H_M(a,b)$ that enters the formula of the finite-$N$ variance of the balanced dispersion in $d=1$ (see Eq. \eqref{eq:1d_order_stat_2ndmomD}). Here, $h(x,y)=(x-y)^2$.
\small{
\begin{align} \label{eq:H_M}
\nonumber H_M(a,b)&=\left(\frac{M}{2}\right)^{\underline{4}} \int _0^a\frac{{\de x}}{a} \int _0^a\frac{{\de y}}{a} \int _0^a\frac{{\de z}}{a} \int _0^a\frac{{\de t}}{a}h(G(x),G(y)) h(G(z),G(t))\\
 \nonumber &+4 \left(\frac{M}{2}\right)^{\underline{3}} \int _0^a\frac{{\de x}}{a} \int _0^a\frac{{\de y}}{a} \int _0^a\frac{{\de z}}{a} \int _0^a\frac{{\de t}}{a}h(G(x),G(y)) h(G(z),G(x))\\
\nonumber &+2 \left(\frac{M}{2}\right)^{\underline{2}} \int _0^a\frac{{\de x}}{a} \int _0^a\frac{{\de y}}{a} \int _0^a\frac{{\de z}}{a} \int _0^a\frac{{\de t}}{a}h(G(x),G(y))h(G(x),G(y)) \\
\nonumber &+4\left(\frac{M}{2}\right)^{\underline{3}}\left(\frac{M}{2}\right)  \int _0^a\frac{{\de x}}{a}\int _0^a\frac{{\de y}}{a}\int _0^a\frac{{\de z}}{a}\int _b^1\frac{{\de t}}{1-b}h(G(x),G(y)) h(G(z),G(t)) \\
\nonumber &+8  \left(\frac{M}{2}\right)^{\underline{2}}\left(\frac{M}{2}\right)  \int _0^a\frac{{\de x}}{a}\int _0^a\frac{{\de y}}{a}\int _0^a\frac{{\de z}}{a}\int _b^1\frac{{\de t}}{1-b}h(G(x),G(y)) h(G(x),G(t)) \\
\nonumber &+4\left(\frac{M}{2}\right)^{\underline{2}}\left(\frac{M}{2}\right)^{\underline{2}}  \int _0^a\frac{{\de x}}{a}\int _b^1\frac{{\de y}}{1-b}\int _0^a\frac{{\de z}}{a}\int _b^1\frac{{\de t}}{1-b}h(G(x),G(y))h(G(z),G(t)) \\
\nonumber &+4 \left(\frac{M}{2}\right)^{\underline{2}}\left(\frac{M}{2}\right)   \int _0^a\frac{{\de x}}{a}\int _b^1\frac{{\de y}}{1-b}\int _0^a\frac{{\de z}}{a}\int _b^1\frac{{\de t}}{1-b} h(G(x),G(y))h(G(x),G(t))\\
\nonumber &+2 \left(\frac{M}{2}\right)^{\underline{2}} \left(\frac{M}{2}\right)^{\underline{2}} \int _0^a\frac{{\de x}}{a}\int _0^a\frac{{\de y}}{a}\int _b^1\frac{{\de z}}{1-b}\int _b^1\frac{{\de t}}{1-b}h(G(x),G(y)) h(G(z),G(t))\\
\nonumber &+4 \left(\frac{M}{2}\right)^{\underline{2}} \left(\frac{M}{2}\right) 
\int _0^a\frac{{\de x}}{a}\int _b^1\frac{{\de y}}{1-b}\int _0^a\frac{{\de z}}{a}\int _b^1\frac{{\de t}}{1-b}
h(G(x),G(y)) h(G(z),G(y))\\
\nonumber &+4\left(\frac{M}{2}\right)^{\underline{2}} 
\int _0^a\frac{{\de x}}{a}\int _b^1\frac{{\de y}}{1-b}\int _0^a\frac{{\de z}}{a}\int _b^1\frac{{\de t}}{1-b}
h(G(x),G(y)) h(G(x),G(y))\\
\nonumber &+8  \left(\frac{M}{2}\right)^{\underline{2}} \left(\frac{M}{2}\right) 
\int _0^a\frac{{\de x}}{a}\int _b^1\frac{{\de y}}{1-b}\int _b^1\frac{{\de z}}{1-b}\int _b^1\frac{{\de t}}{1-b}
h(G(x),G(y))h(G(y),G(t)) \\
\nonumber &+4 \left(\frac{M}{2}\right)^{\underline{3}} \left(\frac{M}{2}\right) 
\int _0^a\frac{{\de x}}{a}\int _b^1\frac{{\de y}}{1-b}\int _b^1\frac{{\de z}}{1-b}\int _b^1\frac{{\de t}}{1-b}
 h(G(x),G(y))h(G(z),G(t))\\
\nonumber &+2 \left(\frac{M}{2}\right)^{\underline{2}} \int _b^1\frac{{\de x}}{1-b}\int _b^1\frac{{\de y}}{1-b}\int _b^1\frac{{\de z}}{1-b}\int _b^1\frac{{\de t}}{1-b}h(G(x),G(y)) h(G(x),G(y))\\
\nonumber &+4 \left(\frac{M}{2}\right)^{\underline{3}} \int _b^1\frac{{\de x}}{1-b}\int _b^1\frac{{\de y}}{1-b}\int _b^1\frac{{\de z}}{1-b}\int _b^1\frac{{\de t}}{1-b}h(G(x),G(y)) h(G(z),G(x))\\
&+\left(\frac{M}{2}\right)^{\underline{4}} \int _b^1\frac{{\de x}}{1-b}\int _b^1\frac{{\de y}}{1-b}\int _b^1\frac{{\de z}}{1-b}\int _b^1\frac{{\de t}}{1-b}h(G(x),G(y)) h(G(z),G(t))\ .
\end{align}
}

\section*{Acknowledgments}
F.D.C. and G.G. are supported by Gruppo Nazionale di Fisica Matematica GNFM-INdAM and by Istituto Nazionale di Fisica Nucleare INFN through the project QUANTUM. F.D.C  acknowledges the support from  PRIN 2022 project 2022TEB52W-PE1-
`The charm of integrability: from nonlinear waves to random matrices', and from PNRR MUR project CN00000013 `Italian National Centre on HPC, Big Data and Quantum Computing'. G.G. acknowledges support from PNRR MUR project PE0000023-NQSTI and from the University of Bari through the 2023-UNBACLE-0245516 grant. 
P.V. acknowledges support from UKRI FLF Scheme (No. MR/X023028/1).

\end{document}